%% file: All.tex
\documentclass[aps,twocolumn,groupedaddress,showpacs]{revtex4-1}

\usepackage{siunitx}
\usepackage{amsmath, amssymb}
\usepackage{color}
\usepackage{nicefrac}
\usepackage{graphicx}
\usepackage{graphics}

\newcommand{\vect}[1]{\boldsymbol{#1}}

\DeclareMathOperator\g{g}
\DeclareMathOperator\sgn{sgn}

\DeclareMathOperator\E{E}

\usepackage{mathtools}
\usepackage{mathrsfs,amsmath}

\usepackage{prettyref}
\usepackage{xr}
\newrefformat{fig}{Fig.~\ref{#1}}
\newrefformat{eq}{Eq.~(\ref{#1})}
\newrefformat{tab}{Tab.~\ref{#1}}

\renewcommand{\vec}[1]{\boldsymbol{#1}}
\begin{document}
\title{Long-lived anomalous thermal diffusion induced by elastic cell membranes on nearby particles}
\author{Abdallah Daddi-Moussa-Ider, Achim Guckenberger and Stephan Gekle}
\affiliation
{Biofluid Simulation and Modeling, Fachbereich Physik, Universit\"at Bayreuth}
\date{\today}





\begin{abstract}
The physical approach of a small particle (virus, medical drug) to the cell membrane represents the crucial first step before active internalization and is governed by thermal diffusion. Using a fully analytical theory we show that the stretching and bending of the elastic membrane by the approaching particle induces a memory in the system which leads to anomalous diffusion, even though the particle is immersed in a purely Newtonian liquid. For typical cell membranes the transient subdiffusive regime extends beyond 10 ms and can enhance residence times and possibly binding rates up to 50\%. Our analytical predictions are validated by numerical simulations.
\end{abstract}
\maketitle

\section{Introduction}

Endocytosis, the uptake of a small particle by a living cell is one of the most important processes in biology \cite{Doherty_2009, Richards_2014, Meinel_2014}. 
Current research is focused mainly on the biophysical and biochemical mechanisms which govern endocytosis when particle and cell are in direct physical contact. 
Much less investigated, yet equally important, is the approach of the particle to the cell membrane before physical contact is established \cite{Junger_2015}. 
In many physiologically relevant situations, e.g., inside the blood stream, the cell and the particle are both suspended in a surrounding liquid and the approach is governed by thermal diffusion of the small particle. The thermal diffusion of small particles (fibrinogen) naturally occurring in human blood has furthermore been suggested as the root cause of red blood cell aggregation \cite{Neu_2002, Steffen_2013, Brust_2014}.

Thermal diffusion of a spherical particle in a bulk fluid is well understood and governed by the celebrated Stokes-Einstein relation. This relation builds a bridge between the particle mobility when an external force is applied to it and the random trajectories observed when only thermal fluctuations are present. Particle mobilities and thermal diffusion near solid walls have been thoroughly investigated 
both theoretically \cite{Goldman_1967, Perkins_1992, Lauga_2005, Felderhof_2005_vacf, Franosch_2009, Yu_2015} 
and experimentally \cite{Banerjee_2005, Holmqvist_2006, Choi_2007, CarbajalTinoco_2007, Schaffer_2007, Huang_2007, Kyoung_2008, Kazoe_2011, Lele_2011, rogers12, lisicki12, Dettmer_2014, lisicki14} finding a reduction of the particle mobility due to the proximity of the wall. 
Some theoretical works have investigated particle mobilities and diffusion close to fluid-fluid interfaces endowed with surface tension \cite{Lee_1980, bickel07, Biawzdziewicz_2010, Bickel_2014} or surface elasticity \cite{Felderhof_2006_elastic, Shlomovitz_2014, Salez_2015} with corresponding experiments \cite{Jeney_2008, Wang_2009_diffusion, Shlomovitz_2013, Zhang_2013_APL, Wang_2014_diffusion, Zhang_2014_plos, Boatwright_2014, saintyves16}. For the case of a membrane with bending resistance transient subdiffusive behavior has been observed in the perpendicular direction \cite{Bickel_2006}. 
Regarding biological cells, recent experiments have measured particle mobilities near different types of cells as well as giant unilamellar vesicles (GUVs), both of which possess an elastic membrane separating two fluids, and found that the mobility near the cell walls does decrease but not as strongly as near a hard wall \cite{Junger_2015}. 

Here we derive a fully analytical theory for the diffusion of a small particle in the vicinity of a realistic cell membrane possessing shear and bending resistance with fluid on both sides. As the typical sizes and velocities are small, the theory is derived in the small Reynolds number regime neglecting the non-linear term, but including the unsteady contribution in the Navier-Stokes equations. Our most important finding is that there exists a long-lasting subdiffusive regime with local exponents as low as 0.87 extending over time scales beyond 10ms. Such behavior is qualitatively different from diffusion near hard walls where the diffusion, albeit being slowed down, still remains normal (i.e. the mean-square-displacement increases linearly with time). Remarkably, our system exhibits subdiffusion in a purely Newtonian liquid whereas most commonly subdiffusion is observed for particles in viscoelastic media. The subdiffusive regime increases the residence time of the particle in the vicinity of the membrane by up to 50\% and is thus expected to be of important physiological significance. Our analytical particle mobilities are quantitatively verified by detailed boundary-integral simulations. Power-spectral densities which are amenable to direct experimental validation using optical traps are provided. 


\section{Results} 
\begin{figure}
\includegraphics[width=\columnwidth]{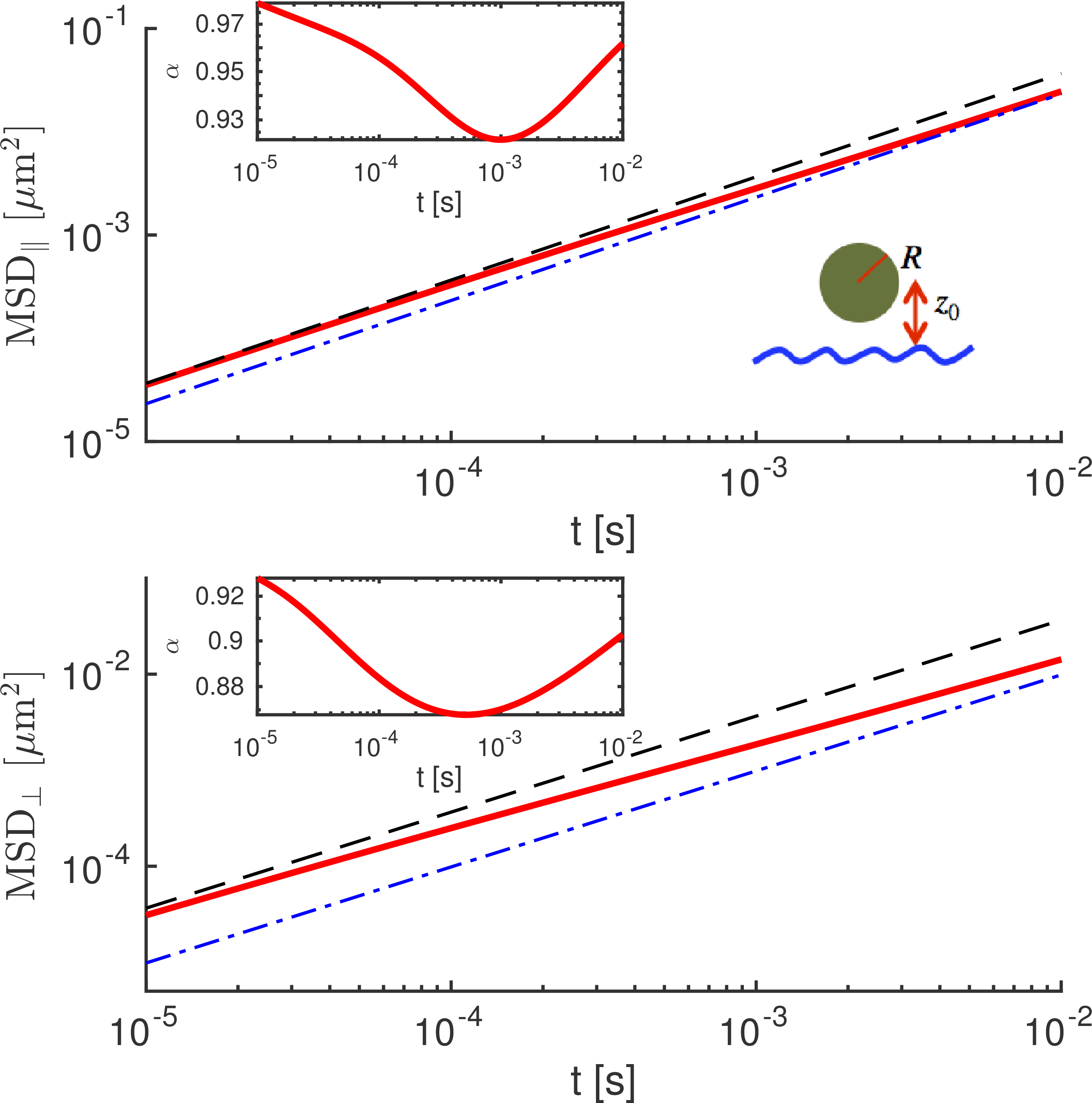}
\caption{Mean-square displacement (red line) of a particle with radius $R$=100nm diffusing $z_0$=153nm above a red-blood cell membrane in lateral (top) and perpendicular (bottom) direction as predicted by our theory at $T=300$K. For short times $t\lesssim 50\mu$s the MSD follows bulk behavior (black dashed line) while for long times the MSD follows hard-wall behavior (blue dash-dotted line). In between, a subdiffusive regime is evident extending up to 10ms and beyond. Insets show the local exponent which goes down until 0.87 for perpendicular diffusion.
\label{fig:MSD} }
\end{figure}

A spherical particle with radius $R=100$nm is located at a distance $z_0=153$nm above an elastic membrane and exhibits diffusive motion as illustrated in the inset of Fig. \ref{fig:MSD}. The membrane has a shear resistance $\kappa_{\mathrm{s}}=5\cdot 10 ^{-6}$N/m and bending modulus $\kappa_{\mathrm{b}}=2\cdot 10 ^{-19}$Nm which are typical values of red blood cells \cite{Freund_2013}. The area dilatation modulus is $\kappa_{\mathrm{a}}=100\kappa_{\mathrm{s}}$. The fluid properties correspond to blood plasma with viscosity $\eta=1.2$mPas. Figure~\ref{fig:MSD} shows the mean-square-displacement (MSD) for parallel as well as perpendicular motion as obtained from our fully analytical theory to be described below. For short times ($t<50\mu$s) the MSD follows a linear behavior with the normal bulk diffusion coefficient $D_0$ since the membrane does not have sufficient time to react on these short scales. This is in agreement with a simple balance between viscosity and elasticity for shear, $\tau_{\mathrm{s}}=\eta R/\kappa_{\mathrm{s}}\approx 37\mu$s, and bending, $\tau_{\mathrm{b}}=\eta R^3/\kappa_{\mathrm{b}}\approx 22\mu$s. For $t>50\mu$s we observe a downward bending of the MSD which is a clear signature of subdiffusive behavior. Indeed, as shown in the insets of Fig. \ref{fig:MSD}, the local exponent $\alpha=\frac{\partial \log \langle x^2 \rangle}{\partial \log t}$ diminishes from 1 down to 0.92 in the parallel and 0.87 in the perpendicular direction. The subdiffusive regime extends up to 10ms in the parallel and even further in the perpendicular direction, which is long enough to be of possible physiological significance. Finally, for long times, the behavior turns back to normal diffusion with $\alpha\approx 1$. Compared to the short-time regime, however, the diffusion coefficient is now significantly lower and approaches the well-known behavior near a solid hard wall with $D_{\mathrm{wall}, \parallel} = D_0(1 - 9/16 R/z_0)$ in the parallel and $D_{\mathrm{wall}, \perp} = D_0(1-9/8R/z_0)$ in the perpendicular case, respectively. Diffusion for long times therefore turns out to depend only on the particle distance and to be independent of the membrane properties.

\begin{figure}
\includegraphics[width=\columnwidth]{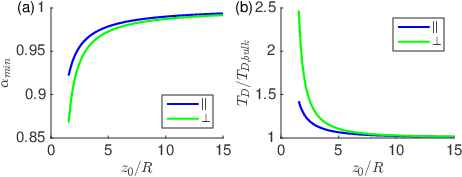}
\caption{(a) Minimum of the local exponent plotted against particle-membrane separation. Significant subdiffusion is observed up to distances roughly ten times the particle radius. (b) The time required to diffuse one particle radius increases due to the presence of the membrane thus leading to an enhanced residence time of the particle in the vicinity of the membrane which may increase the probability of triggering endocytosis.
\label{fig:distance} }
\end{figure}

In Fig. \ref{fig:distance}~(a) we show the minimum of the local exponent for different particle-membrane separations. Even for distances ten times the particle radius, a significant deviation of the local exponent from 1 is still observable. From the MSDs it is straightforward to estimate the time $T_D$ required by the particle to diffuse a distance equal to its own radius which gives an approximate measure of the ''diffusion speed''. As expected based on the data from Fig. \ref{fig:MSD}, diffusion in the perpendicular direction is slowed down significantly more than for lateral motion, see Fig. \ref{fig:distance}~(b), in agreement with recent experimental observations \cite{Junger_2015}.

\begin{figure}
\includegraphics[width=\columnwidth]{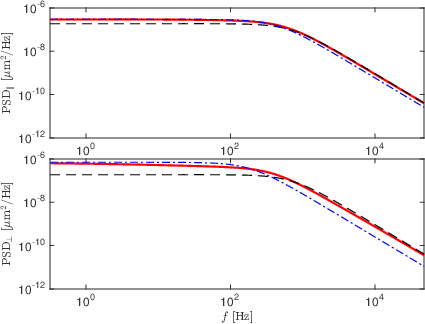}
\caption{Predicted power-spectral density of position fluctuations if the same bead as in Fig. \ref{fig:MSD} is confined by a typical optical trap of strength $K=10^{-5}$N/m \cite{Junger_2015}. Similar as in the MSD of Fig. \ref{fig:MSD} a transition from hard-wall-like behavior (blue dash-dotted line) for low frequencies to bulk-like behavior (black dashed line) at high frequencies is seen. 
\label{fig:PSD} }
\end{figure}

Experimentally, long MSDs can be difficult to measure as the particle may move out of the focal plane during the recording time. A commonly used technique is therefore to confine the particle to its position using optical traps. One then records the power spectral density (PSD) of particle fluctuations around its equilibrium position. The PSDs predicted by our theory for a typical optical trap with spring constant $K=10^{-5}$N/m \cite{Junger_2015} as a function of frequency $f=\omega/2\pi$ are shown in Fig. \ref{fig:PSD}. The general behavior of the unconstrained system is not qualitatively altered by the optical confinement: for high frequencies the behavior is bulk-like (mirroring the bulk-like MSD at short times) while for low frequencies the PSD approaches that expected near a solid wall (mirroring the hard-wall like MSD at long times). The frequency range of the transition lies mainly below 1 kHz and should thus be experimentally accessible.


\section{Theory}

Our theoretical development leading to figures~\ref{fig:MSD} through \ref{fig:PSD} proceeds via the calculation of particle mobilities and the fluctuation-dissipation theorem and can be sketched as follows (a detailed derivation is given in Appendices \ref{AppendixMembraneMechanics}-\ref{sec:diffusion}). We consider a spherical particle of radius $R$ driven by an oscillating force $\vec{F}_\omega(t)=\vec{F}_0e^{i\omega t}$ in a fluid with density $\rho$ and dynamic viscosity $\eta$ whose complex mobility $\mu(\omega)$ for a fixed $\omega$ is defined as
\begin{equation}
\vec{V}_\omega\left(t\right) = \mu\left(\omega\right) \vec{F}_\omega\left(t\right)
\label{eqn:defMobility}
\end{equation}
and can be separated into the three contributions
\begin{equation}
\mu\left(\omega\right) = \mu_0 + \mu_0^u(\omega) + \Delta\mu\left(\omega, z_0\right).
\end{equation}
Here, $\mu_0=1/(6\pi\eta R)$ is the usual steady-state bulk mobility, 
\begin{equation}
\mu_0^u(\omega)=\mu_0\left(e^{-R\lambda\sqrt{-i}} - 1\right)
\end{equation}
with $\lambda^2=\rho\omega/\eta$ is the correction due to fluid inertia \cite{pozrikidis11} and $\Delta\mu\left(\omega, z_0\right)$ is the correction due to the elastic membrane at distance $z_0$. In order to derive the mobility corrections, we employ the commonly used approximation of a small particle ($R/z_0\ll 1$). Using numerical simulations of a truly extended particle, we will show below that this approximation is surprisingly good even for $R/z_0=0.65$. The problem is thus equivalent to solving the unsteady Stokes equations with an arbitrary time dependent point force $\vect{F}$ located at $\vect{r}_0$
\begin{eqnarray}
-\rho\frac{\partial \vect{v}}{\partial t}+\eta\vec{\nabla}^2\vec{v} - \vec{\nabla}p + \vect{F}\delta(\vect{r} - \vect{r}_0)  &=& 0\notag\\
\vec{\nabla}\cdot\vec{v} & = & 0
\label{eqn:Stokes}
\end{eqnarray}
with the fluid velocity $\vec{v}$, the pressure $p$ and the point force position $\vect{r}_0$. The elastic membrane is located at $z=0$, has infinite extent in $x$ and $y$ directions and is surrounded by fluid on both sides. Following the usual approximation of small deformations, we impose the traction jump at $z=0$ which follows from the Skalak \cite{skalak73} and Helfrich \cite{helfrich73} laws for the shear and bending resistance as detailed in Appendix~\ref{AppendixMembraneMechanics}
\begin{eqnarray}
\Delta f^x & = & -\frac{\kappa_{\mathrm{s}}}{3}\left(2\left(1+C\right)u_{x,xx} + u_{x,yy} + \left(1+2C\right)u_{y,xy}\right)    \notag \\
\Delta f^y & = & -\frac{\kappa_{\mathrm{s}}}{3}\left( u_{y,xx}+2\left(1+C\right)u_{y,yy} + \left(1+2C\right)u_{x,xy}\right) \notag \\
\Delta f^z & = & \kappa_{\mathrm{b}}\left(u_{z,xxxx} + 2u_{z,xxyy} + u_{z, yyyy}\right)
\label{eqn:stressJump}
\end{eqnarray}
where the membrane deformation is $\vec{u}$ and the notation $u_{,\cdot}$ denotes partial spatial derivatives. The moduli are $\kappa_{\mathrm{s}}$ for shear resistance and $\kappa_{\mathrm{b}}$ for bending resistance while the ratio between shear and area dilatation modulus is $C=\kappa_{\mathrm{a}}/\kappa_{\mathrm{s}}$. The no-slip condition at the membrane surface relates the surface deformation to the local fluid velocity
\begin{equation}
\frac{d\vec{u}}{dt} = \vec{v}|_{z=0}.
\label{eqn:noSlip}
\end{equation}
Together with equations~(\ref{eqn:Stokes}), (\ref{eqn:stressJump}) and (\ref{eqn:noSlip}) this represents a closed mathematical problem for the velocity field $\vec{v}$. For its solution, the Stokes equations (\ref{eqn:Stokes}) are first Fourier-transformed into frequency space. The dependency on the $x$ and $y$ coordinates is Fourier-transformed into wave vectors $q_x$ and $q_y$ which subsequently allows us to consider the contributions of the longitudinal and transversal velocity components separately \cite{bickel07}. After eliminating the pressure, this leads to three differential equations for the three velocity components for which an analytical solution can be found. From the velocity field the mobility correction of the particle is directly obtained. The details are given in Appendix~\ref{sec:mobilities}.

The mobility correction is a tensorial quantity which in the present case has two components for the mobility parallel $\Delta\mu_\parallel(\omega, z_0)$ and perpendicular $\Delta\mu_\perp(\omega, z_0)$ to the membrane. 
Furthermore, the mobility correction in each direction can be split into a contribution $\Delta\mu_{\mathrm{b}}$ due to bending resistance and a contribution $\Delta\mu_{\mathrm{s}}$ due to shear resistance and area dilatation. The final results  are conveniently expressed in terms of the dimensionless numbers:
\begin{eqnarray}
\beta &=& \frac{12z_0\eta\omega}{\kappa_{\mathrm{s}}+\kappa_{\mathrm{a}}} \, ,\notag\\
\beta_{\mathrm{b}} &=& 2z_0\left(\frac{4\eta\omega}{\kappa_{\mathrm{b}}}\right)^{1/3} \, ,\notag\\
\sigma &=&z_0 \left(\frac{\rho\omega}{\eta}\right)^{1/2} \, ,
\end{eqnarray}
where $\beta$ captures the effect of shear resistance and area dilatation, $\beta_\mathrm{b}$ the effect of bending resistance and $\sigma$ the effect of fluid inertia on the mobility corrections. 

The mobility corrections are
\begin{eqnarray}
\frac{ \Delta \mu_{\parallel,\mathrm{s}}}{\mu_0} &=& \frac{3i}{\sigma^2} \frac{R}{z_0} 
 \int_0^{\infty} \frac{s^3 (re^{-r} - se^{-s})^2}{4 (r-s)s^2 - \beta \sigma^2} d  s \nonumber\\
 &+&
 \frac{3i}{4} \frac{R}{z_0} \int_0^{\infty} \frac{s^3 e^{-2r}}{r\left( \frac{1+C}{2} \beta r  -i s^2 \right)} d  s,
\label{eqn:mu_para_s_unsteady}\\
\frac{\Delta \mu_{\parallel, \mathrm{b}}}{\mu_0} &=& \frac{3i}{\sigma^2} \frac{R}{z_0} \int_0^{\infty} \frac{4rs^7 (e^{-r}-e^{-s})^2}{16s^5 (r-s) - r\beta_\mathrm{b}^3 \sigma^2} d  s, \\
 \frac{\Delta \mu_{\perp, \mathrm{s}} }{\mu_0} &=& \frac{6i}{\sigma^2} \frac{R}{z_0} \int_0^{\infty}   \frac{s^5 \left( e^{-s}-e^{-r} \right)^2}
 {  4(r-s)s^2-\sigma^2 \beta} d s,\label{eqn:mu_perp_s_unsteady}\\
  \frac{\Delta \mu_{\perp,\mathrm{b}}}{\mu_0} &=& \frac{6i}{\sigma^2} \frac{R}{z_0} \int_0^{\infty} \frac{4s^7(s e^{-r}-r e^{-s})^2}{r(16 s^5(r-s)-r \beta_\mathrm{b}^3 \sigma^2)} d  s,
\label{eqn:mu_perp_b_unsteady}
\end{eqnarray}
with $r=\sqrt{s^2+i\sigma^2}$. The integrals are well-behaved and thus amenable to straightforward numerical integration. 
The effect of inertia on the diffusion has recently been investigated in bulk systems \cite{Franosch_2011, Li_2013_AnnPhys, Duplat_2013, Kheifets_2014, Pesce_2014}.
However, as shown in the Supporting Information
\footnote{See Supplemental  Material  at [URL will be inserted by publisher] for a comparison between the steady and the unsteady mobility corrections for the physical parameters corresponding to Fig. \ref{fig:MSD} with a fluid density $\rho = 10^3$ kg/m$^3$.},
for the realistic situation treated in figure~\ref{fig:MSD}, the contribution of fluid inertia is completely negligible in the frequency range that is affected by membrane elasticity which is the focus of this work.

In the following, we will thus consider the case $\sigma=0$, for which an analytical solution is possible:

\begin{widetext}
\begin{eqnarray} 
\frac{\Delta \mu_{\parallel,\mathrm{s}} }{\mu_0} &=& \frac{3}{8}\frac{R}{z_0} \Bigg( -\frac{5}{4} + \frac{\beta^2}{8}-\frac{3i\beta}{8} + i\beta(1+C) e^{i\beta(1+C) } \E_1(i\beta(1+C))
 +\left(-\frac{\beta^2}{2} + \frac{i\beta}{2} \left(1-\frac{\beta^2}{4} \right) \right)e^{i\beta} \E_1(i\beta)
 \Bigg),\label{eqn:paraShear}
 \\
 \frac{\Delta \mu_{\parallel, \mathrm{b}} }{\mu_0} &=& \frac{3}{64} \frac{R}{z_0} \left( -2 + \frac{i\beta_{\mathrm{b}}^3}{3}\left( \phi_+ + e^{-i\beta_{\mathrm{b}}}\E_1(-i\beta_{\mathrm{b}}) \right) \right),\label{eqn:paraBen}
 \\
 \frac{\Delta {\mu}_{\perp, \mathrm{s}} }{\mu_0} &=& 
 \frac{9i}{16} \frac{R}{z_0} \frac{1}{\beta} \left( 1-4 e^{i \beta} \E_5(i \beta) \right),\label{eqn:perpShear}
 \\
 \frac{\Delta {\mu}_{\perp, \mathrm{b}}}{\mu_0} &=& 
\frac{3i \beta_{\mathrm{b}}}{8} \frac{R}{z_0}
  \Bigg(
 \left( \frac{\beta_{\mathrm{b}}^2}{12}+\frac{i\beta_{\mathrm{b}}}{6} + \frac{1}{6}\right)\phi_+
 + \frac{\sqrt{3}}{6} (\beta_{\mathrm{b}}+i)\phi_-
+  \frac{5i}{2\beta_{\mathrm{b}}} + e^{-i\beta_{\mathrm{b}}} \E_1(-i\beta_{\mathrm{b}}) \left( \frac{\beta_{\mathrm{b}}^2}{12} -\frac{i\beta_{\mathrm{b}}}{3} -\frac{1}{3}\right)
 \Bigg),\label{eqn:perpBen}
\end{eqnarray}
\end{widetext}
with 
\begin{equation}
 \phi_{\pm} =  e^{-i \overline{z_{\mathrm{b}}}} \E_1 \left(-i\overline{z_{\mathrm{b}}} \right) \pm  e^{-i z_{\mathrm{b}}} \E_1 \left(-iz_{\mathrm{b}} \right)
\end{equation}
where $z_{\mathrm{b}}=j\beta_\mathrm{b}$ and $j=e^{2i\pi/3}$. Bar denotes complex conjugate. $\E_n$ denotes the exponential integral $\E_n(x)=\int_1^\infty e^{-xt}/t^n dt$ \cite{Abramowitz_book}.

From the frequency-dependent mobilities the mean-square displacement in a thermally fluctuating system can be computed using the fluctuation-dissipation theorem with the velocity autocorrelation function $\phi_{v} (t)$ as an intermediate step \cite{kubo85} as detailed in Appendix~\ref{sec:diffusion}
\begin{eqnarray}
\phi_{v} (t)&=&\frac{k_{\mathrm{B}} T}{2\pi}\int_{-\infty}^\infty \mu(\omega)e^{i\omega t}d\omega \label{VACFDefinition}\\
\left<x(t)^2\right>&=&2\int_0^t (t-s)\phi_v(s)ds \label{eqn:MSD_from_mu}.
\end{eqnarray}
Using the mobilities from Eqs.~(\ref{eqn:paraShear}) - (\ref{eqn:perpBen}), the MSD can be analytically computed and the resulting equations are given in Appendix~\ref{sec:diffusion}. In order to compute the MSDs shown in figure~\ref{fig:MSD} mobilities are calculated using the initial particle-membrane distance $z_0$,
which is equivalent to assuming a not too large deviation of the particle from its initial position.

Similarly, the power spectral densities of the position fluctuations as shown in figure~\ref{fig:PSD} can be calculated as \cite{Franosch_2009}
\begin{equation}
S(\omega) = \frac{2k_{\mathrm{B}}T \mathrm{Re}\left[\mu(\omega)^{-1}\right]}{\left( \omega \mathrm{Re}\left[ \mu(\omega)^{-1}\right]\right)^2 + 
                     \left(\omega \mathrm{Im}\left[\mu(\omega)^{-1}\right] + K\right)^2}.
\end{equation}


\section{Mobility Simulations} 

\begin{figure}
\includegraphics[width=\columnwidth]{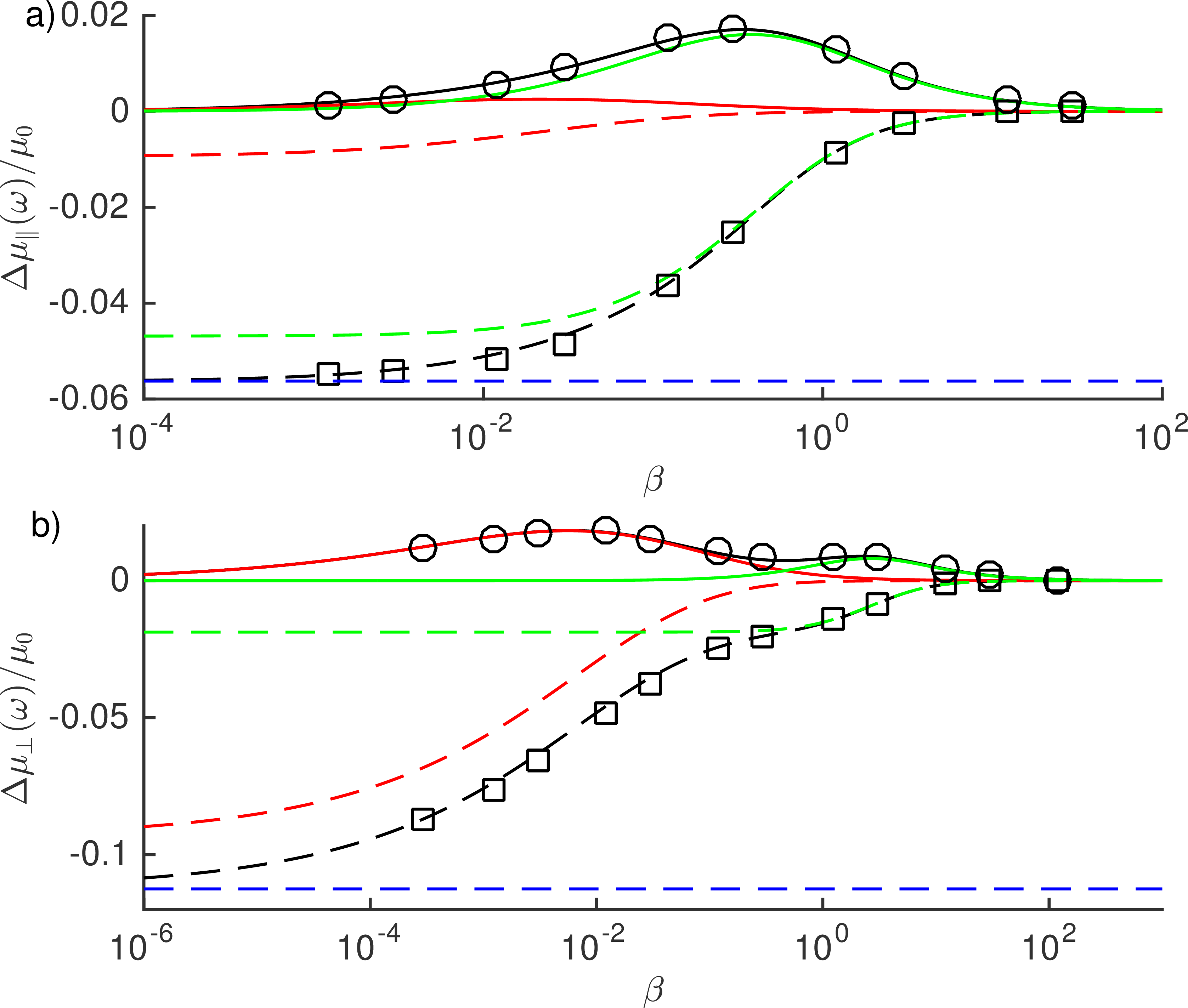}
\caption{The complex mobility of a spherical particle driven by a sinusoidal force with frequency $\omega$ situated a distance $z_0$ above the membrane. Theoretical predictions from Eqs.~\eqref{eqn:paraShear}-\eqref{eqn:perpBen} are shown as black dashed lines (real part) and black solid lines (imaginary part) and compared to BIM simulations shown as circles (real part) and squares (imaginary part). The green and red lines show the contributions due to shear and bending resistance, respectively. 
 For $R/z_0=0.1$ (with $C=1$, $\kappa_\mathrm{s}R^2/\kappa_\mathrm{b}=2$) the agreement between theory and simulations is excellent. 
 For very low frequencies the hard wall behavior is obtained (blue dashed line). 
\label{mobility_artificial} }
\end{figure}

We use boundary-integral (BIM) simulations to obtain a direct validation of the frequency-dependent mobilities and to assess the accuracy of the point-particle approximation for finite-radius particles. BIMs are a standard method for solving the steady Stokes equations \cite{Pozrikidis_book} including elastic surfaces \cite{Pozrikidis_2001}. Some details on our implementation are given in the SI. Compared with most other flow solvers, BIMs have the advantage that they are able to treat a truly inifinite fluid domain thus excluding artifacts due to periodic replications of the system. 

We simulate a spherical particle driven by an oscillating force with frequency $\omega$. By recording the instantaneous particle velocity, the mobility correction $\Delta\mu(\omega)$ can be obtained from the amplitude ratio and the phase shift between force and velocity as illustrated in the SI.

\begin{figure}
\includegraphics[width=\columnwidth]{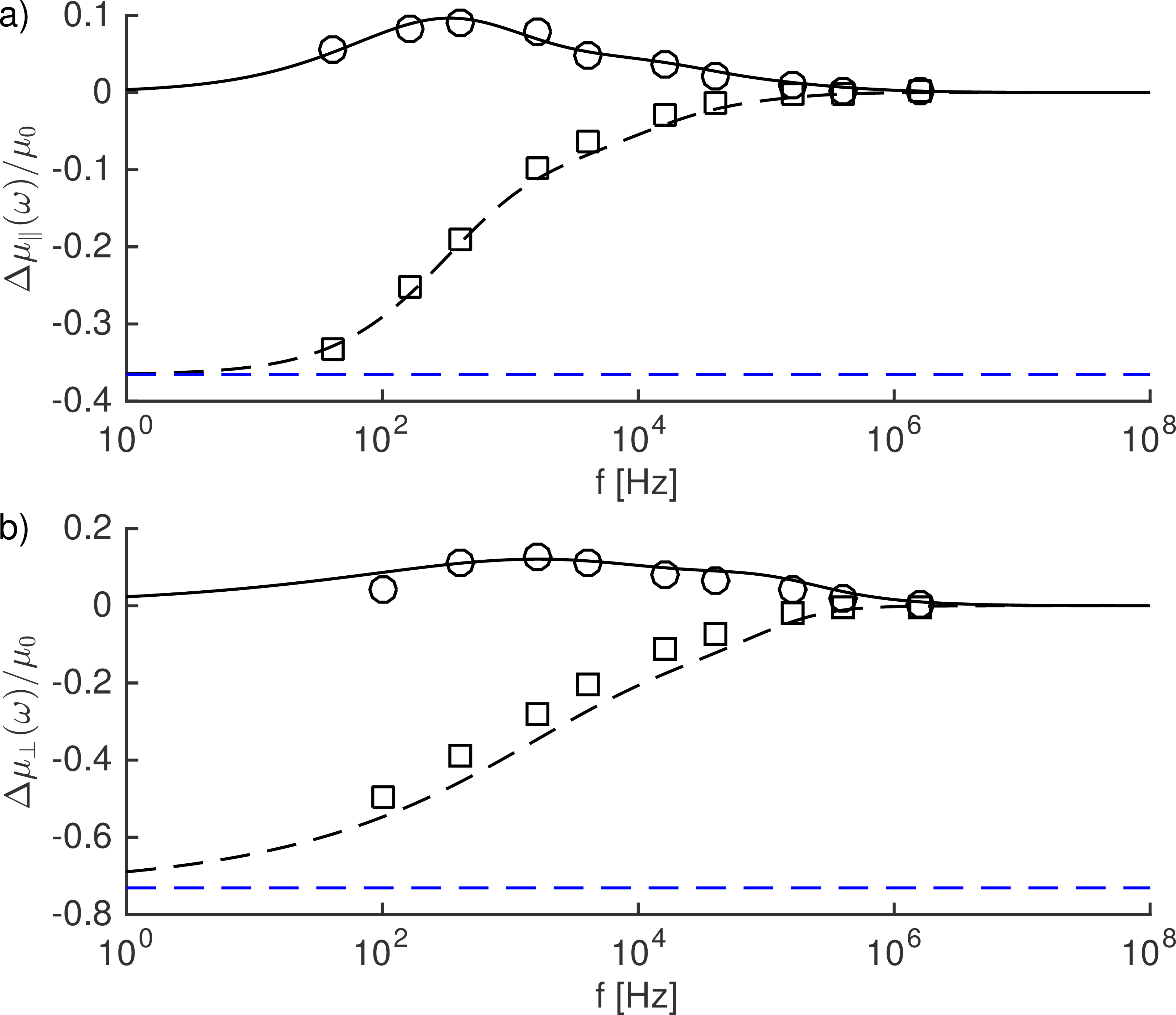}
\caption{
The real part (dashed lines) and the imaginary part (solid lines) of the  complex mobility for a particle moving parallel (a) or perpendicular (b) to a realistically modeled red blood cell membrane with parameters corresponding to Fig. \ref{fig:MSD}. Even for $R/z_0=0.65$ as used here the agreement is still good. 
\label{mobility_diff} }
\end{figure}

In Fig.~\ref{mobility_artificial} we compare our theoretical prediction to the result of BIM simulations with $R/z_0=0.1$ and find excellent agreement. Splitting the mobility correction into the contributions due to shear/area resistance (green line in Fig. \ref{mobility_artificial}) on the one hand and bending resistance (red line) on the other, we find that bending resistance manifests itself at significantly lower frequencies than shear resistance. As might intuitively be expected, the parallel mobility is mainly determined by shear resistance, while for the perpendicular mobility bending resistance dominates. Yet, we note that for both directions, shear/area resistance and bending resistance are important. This becomes apparent especially at low frequencies: neither shear/area resistance nor bending resistance alone are able to recover the hard-wall limit. As shown in the SI, a similar effect appears in the limit of infinitely stiff membranes: only if shear and bending stiffness both tend to infinity does one recover the hard-wall limit.

Finally, we investigate the validity of the point-particle approximation for particles close to the interface. For this, we use the parameters as in Fig. \ref{fig:MSD}. Even for $R/z_0=0.65$ the agreement is still surprisingly good as shown in Fig. \ref{mobility_diff}.


\section{Conclusion} 
We have presented a fully analytical theory for the thermal diffusion of a small spherical particle in close vicinity to an elastic cell membrane. The frequency-dependent particle mobilities predicted by the theory are in excellent agreement with boundary-integral simulations, even for surprisingly large particles where the point-force approximation made in the theory is no longer strictly valid. Independent of the membrane properties, the mean-square displacement is shown to be bulk-like at short and hard-wall-like at very long times. In between, however, there exists a significant time span during which the particle shows subdiffusion with exponents as low as 0.87. For membrane parameters corresponding to a typical red blood cell the subdiffusive regime extends up to and beyond 10ms and may thus be of possible physiological significance, e.g., for the uptake of drug carriers or viruses by a living cell. Our results can be directly verified experimentally by comparing the power-spectral densities of the position fluctuations in Fig. \ref{fig:PSD}.

In living cells the membrane elastic properties depend on the local cholesterol level \cite{Byfield_2004} which can lead to localized patches of varying stiffness. According to our calculations, adjusting the shear/bending rigidity would allow the cell to specifically influence the endocytosis probability: An enhanced bending stiffness combined with reduced shear elasticity would reduce perpendicular diffusion -- keeping the approaching particle close to the membrane for a longer time -- and at the same time enhance parallel diffusion -- allowing the particle to survey more quickly the cell surface for favorable biochemical binding sites. 

\begin{acknowledgments}
The authors gratefully acknowledge funding from the Volkswagen Foundation and the KONWIHR network as well as computing time granted by the Leibniz-Rechenzentrum on SuperMUC.
\end{acknowledgments}



\begin{appendix}

\section{MEMBRANE MECHANICS}\label{AppendixMembraneMechanics}

In this appendix, we give the derivation of the linearized tangential and normal traction jumps as stated in Eq.~\eqref{eqn:stressJump} of the main text. Initially, the interface is described by the infinite plane $z=0$. Let the position vector of a material point  before deformation be $\vect{A}$, and $\vect{a}$ after deformation. In the undeformed state, we have $\vect{A}(x, y) =  x \vect{e}_x + y \vect{e}_y$, where $\vect{e}_i$, with $i \in \{x, y, z\}$ are the Cartesian base vectors. Hereafter, we shall reserve the capital roman letters for the undeformed state. The membrane can be defined using two covariant base vectors $\vect{a}_1$ and $\vect{a}_2$, together with the normal vector $\vect{n}$.
$\vect{a}_1$ and $\vect{a}_2$ are the local non-unit tangent vectors to coordinate lines. 
In the Cartesian coordinate system, $\vect{a}_1 =  \vect{a}_{,x}$ and $\vect{a}_2 =  \vect{a}_{,y}$, where the comma denotes a spatial derivative.
The unit normal vector to the interface reads
\begin{equation}
 \vect{n} = \frac{\vect{a}_1 \times \vect{a}_2}{|\vect{a}_1 \times \vect{a}_2|}.
\end{equation}
It can be seen that the covariant base vectors in the undeformed state are identical to those of the Cartesian base. 
The displacement vector of a point on the membrane can be written as
\begin{equation}
 \vect{u} = \vect{a} - \vect{A} = u_x \vect{e}_x + u_y \vect{e}_y + u_z \vect{e}_z.
\end{equation}

The covariant base vectors are therefore
\begin{eqnarray}
 \vect{a}_1 &=& \left(1+u_{x,x}\right)\vect{e}_x + u_{y,x} \vect{e}_y + u_{z,x} \vect{e}_z,\label{a1}\\
 \vect{a}_2 &=& u_{x,y} \vect{e}_x + \left(1+u_{y,y}\right) \vect{e}_y + u_{z,y} \vect{e}_z,\label{a2}
\end{eqnarray}
and the linearized normal vector reads
\begin{equation}
 \vect{n} \approx -u_{z,x} \vect{e}_x - u_{z,y} \vect{e}_y + \vect{e}_z.
\end{equation}

The components of the metric tensor in the deformed state are defined by the inner product $a_{\alpha \beta} = \vect{a}_\alpha.\vect{a}_\beta$. 
Note that $A_{\alpha \beta}$ is then nothing but the second order identity tensor $\delta_{\alpha \beta}$. 
From Eqs. \eqref{a1} and \eqref{a2}, $a_{\alpha\beta}$ can straightforwardly be computed.
The contravariant tensor (conjugate metric) is the inverse of the covariant tensor.
We directly have to the first order
\begin{equation}
a^{\alpha \beta} \approx
 \left( \begin{array}{cc}
        1 - 2 u_{x,x} & -2\epsilon\\
        -2\epsilon & 1 - 2 u_{y,y}
        \end{array}
\right),
\label{contravariantMetricTensor}
\end{equation}
where $2\epsilon = u_{x,y}+u_{y,x}$ is the engineering shear strain.
In the following, we will first derive the in-plane stress tensor. A resistance to bending will be added independently by assuming a linear isotropic model equivalent to the Helfrich model for small deformations \cite{pozrikidis01jfm}. 


\subsubsection{In-plane stress tensor}

Here we use the Einstein summation convention, in which a covariant index followed by the identical contravariant index (or vice versa) is implicitly summed over the index. The two invariants of the transformation are given by Green and Adkins  \cite{green60}
\begin{eqnarray}
 I_1 &=&  A^{\alpha \beta} a_{\alpha \beta} - 2,\\
 I_2 &=& \operatorname{det}A^{\alpha \beta} \operatorname{det}a_{\alpha \beta} - 1,
\end{eqnarray}
where $A^{\alpha \beta} = \delta_{\alpha \beta}$, is the contravariant metric tensor of the undeformed state.
The two invariants are found to be  equal and they are given by $I_1 = I_2 = 2 e = 2 \left( u_{x,x} + u_{y,y} \right)$, where $e$ denotes the dilatation.
The contravariant components of the stress tensor $\tau^{\alpha \beta}$ are related to the strain energy $W(I_1, I_2)$ via a constitutive law. We have \cite{lac04}
\begin{equation}
 \tau^{\alpha \beta} = \frac{2}{J_s} \frac{\partial W}{\partial I_1} A^{\alpha \beta} + 2 J_s \frac{\partial W}{\partial I_2} a^{\alpha \beta},
 \label{stressTensor}
\end{equation}
where $J_s \coloneqq \sqrt{1+I_2} \approx 1+ e$ is  the Jacobian determinant, representing the ratio between the deformed and undeformed local surface area. 

Several models have been proposed in order to describe the mechanics of elastic membranes. The neo-Hookean model is characterized by a single parameter containing the membrane elastic shear and area dilatation modulus, while the Skalak model \cite{skalak73} uses two separate parameters for shear and area dilatation resistance, respectively. The strain energy in the Skalak model reads \cite{kruger11}
\begin{equation}
 W^{\mathrm{SK}} = \frac{\kappa_\mathrm{s}}{12} \left(  (I_1^2 + 2I_1 - 2I_2) +C I_2^2 \right),
 \label{W_SK}
\end{equation}
where $C = \kappa_\mathrm{a}/\kappa_\mathrm{s}$ is the ratio between area dilation and shear modulus. By taking $C = 1$, the Skalak model predicts the same behavior as the neo-Hookean for small deformations \cite{lac04}. 
The calculations yield to the first order a stress tensor in the form of
\begin{equation}
\tau^{\alpha \beta} \approx 
   \frac{2\kappa_\mathrm{s}}{3}
 \left( \begin{array}{cc}
         u_{x,x}+Ce & \epsilon \\
          \epsilon & u_{y,y}+Ce
        \end{array}
\right).
\label{stressTensorSkalakGeneral}
\end{equation}


\subsubsection{Bending resistance}

Under the action of an external load, the initially plane membrane bends. For small membrane curvatures, the bending moment $M$ can be related to the curvature tensor via the linear isotropic model  \cite{pozrikidis01jfm,zhao10}
\begin{equation}
 M_\alpha^\beta = -\kappa_\mathrm{b} (b_\alpha^\beta - B_\alpha^\beta),
\end{equation}
where $\kappa_\mathrm{b}$ is the bending modulus, having the dimension of energy. Here $b_\beta^\alpha$ is the mixed version of the second fundamental form which follows from the curvature tensor (second fundamental form)
\begin{equation}
b_{\alpha \beta} = \vect{n} .  \vect{a}_{\alpha,\beta} \;\;\; \mathrm{for} \;\;\;\alpha, \beta \in \{1,2\}
\end{equation}
via $b_\alpha^\beta = b_{\alpha \delta} a^{\delta \beta} \approx  u_{z, \alpha \beta}$.
As the surface reference is a flat membrane, $B_{\alpha}^{\beta}$ therefore vanishes.
The bending moment reads
\begin{equation}
 M_{\alpha}^{\beta} \approx -  \kappa_\mathrm{b}  u_{z, \alpha \beta}.
 \label{mixedMomentTensor}
\end{equation}
The surface transverse shear vector $\vect{Q}$ is obtained from a local torque balance with the exerted moment by \cite{zhao10}
\begin{equation}
 \nabla_{\alpha} M^{\alpha \beta} - Q^\beta = 0,
\end{equation}
where $\nabla_\alpha$ is the covariant derivative defined for a contravariant tensor $M^{\alpha \beta}$ by 
\begin{equation}
 \nabla_\lambda M^{\alpha \beta} = \partial_\lambda M^{\alpha \beta} + \Gamma_{\lambda \eta}^\alpha   M^{\eta \beta} + \Gamma_{\lambda \eta}^\beta  M^{\alpha \eta},
 \label{covariantDerivative}
\end{equation}
where $\Gamma_{\alpha \beta}^\lambda$ are the Christoffel symbols of the second kind, defined by $\Gamma_{\alpha \beta}^\lambda= \vect{a}_{\alpha,\beta}.\vect{a}^\lambda$, and $\vect{a}^{\lambda}$ are the contravariant basis vectors, which are related to those of the covariant basis via the contravariant metric tensor by $\vect{a}^{\alpha} = a^{\alpha \beta} \vect{a}_\beta$.
To first order, only the partial derivative in Eq. \eqref{covariantDerivative} remains.

The raising and lowering indices operation on the second order tensor $M$ implies that $M^{\alpha \beta} = a^{\alpha \gamma} a^{\beta \delta} M_{\gamma \delta}$,
which, to the first order, is the same as $M_{\alpha}^{\beta}$ given by Eq. \eqref{mixedMomentTensor}.
The contravariant component of the transverse shear vector is therefore
\begin{equation}
 Q^{\beta} \approx - \kappa_\mathrm{b} u_{z, \alpha \beta \alpha}.
\end{equation}


\subsubsection{Equilibrium Equation}

The membrane equilibrium condition including both the shear and the bending forces reads \cite{zhao10}
\begin{eqnarray}
 \nabla_\alpha {\tau^{\alpha \beta}} - b_{\alpha}^{\beta} Q^{\alpha} &=& -{\Delta f^{\beta}} \label{tangentialConditionIncludingBending},\\
 \tau^{\alpha \beta} b_{\alpha \beta} + \nabla_\alpha Q^\alpha &=& -\Delta f^{z},\label{normalConditionIncludingBending} 
\end{eqnarray}
where $\Delta f^{\beta}$, with $\beta \in \{x,y\}$ is the tangential traction jump  at the elastic wall, and $\Delta f^{z}$ is the normal traction jump.
The second term on the left-hand side (LHS) of Eq. \eqref{tangentialConditionIncludingBending} is  irrelevant in the first order approximation.
The same is true for the first term on the LHS of Eq. \eqref{normalConditionIncludingBending}.

Finally, the linearized traction jump across the membrane is
 \begin{eqnarray}
\frac{\kappa_\mathrm{s}}{3} \big( \Delta_{\parallel} u_\beta + (1+2C) e_{,\beta} \big) &=& -\Delta f^{\beta},\notag\\
 \kappa_\mathrm{b}   \Delta_{\parallel}^2 u_z &=& +{\Delta f^{z}},
 \label{eqn:LaplaceBeltrami}
\end{eqnarray}
where  $\Delta_\parallel f = f_{,xx}+f_{,yy}$ is the horizontal Laplace-Beltrami of a given function $f$. 
Eqs.~(\ref{eqn:LaplaceBeltrami}) are equivalent to Eqs.~(\ref{eqn:stressJump}) of the main text.


\section{DERIVATION OF PARTICLE MOBILITIES}
\label{sec:mobilities}

\subsection{Hydrodynamic equations in Fourier space}

We start by transforming Eqs.~(\ref{eqn:Stokes}) of the main text to Fourier space. The spatial 2D Fourier transform for a given function $f$ is defined as
\begin{equation}
 \mathscr{F} \{ f(\boldsymbol{\rho}) \} = \tilde{f}(\vect{q}) = 
 \int_{\mathbb{R}^2} f(\boldsymbol{\rho}) e^{-i \vect{q} . \boldsymbol{\rho}} d ^2 \boldsymbol{\rho},
 \label{2DFourierTransform}
\end{equation}
where  $\boldsymbol{\rho} = (x,y)$ is the projection of the position vector $\vect{r}$ onto the horizontal plane, and $\vect{q} = (q_x, q_y)$ is the Fourier transform variable. Similarly as in Bickel \cite{bickel07}, all the vector fields are subsequently decomposed into longitudinal, transversal and normal components. For a given quantity $\tilde{\vect{Q}}$, whose components are $(\tilde{Q}_x, \tilde{Q}_y)$ in the Cartesian coordinate base, its components in the new orthogonal base $(\tilde{Q}_l, \tilde{Q}_t)$ are given by the following transformation
\begin{equation}
\left( 
	\begin{array}{c}
         \tilde{Q}_x \\
         \tilde{Q}_y 
        \end{array}
\right)
=
\frac{1}{q}
 \left( 
	\begin{array}{cc}
         q_x & q_y\\
         q_y & -q_x
        \end{array}
\right)
\left( 
	\begin{array}{c}
         \tilde{Q}_l \\
         \tilde{Q}_t 
        \end{array}
\right),
\label{transformation}
\end{equation}
where $q \coloneqq |\vect{q}|$.
Note that the inverse transformation is given also by Eq. \eqref{transformation}.
Since the membrane shape depends on the history of the particle motion we also perform a Fourier analysis in time which for a function $f(t)$ is 
\begin{equation}
 \mathscr{F}\{f(t)\} = f(\omega) = \int_{\mathbb{R}} f(t) e^{-i \omega t} d  t.
\end{equation}
In the following, the Fourier-transformed function pair $f(t)$ and $f(\omega)$ are distinguished only by their argument while the tilde is reserved to denote the spatial 2D Fourier transforms. The unsteady Stokes equations~(\ref{eqn:Stokes}) thus become
\begin{eqnarray}
 -(i\rho \omega + \eta q^2)\tilde{v}_l + \eta \tilde{v_l}_{,zz} - iq \tilde{p} +F_l \delta (z-z_0) &=& 0~~\label{longitudinalStokesEqn}\\
 -(i\rho \omega + \eta q^2)\tilde{v}_t + \eta \tilde{v_t}_{,zz}  +F_t \delta (z-z_0) &=& 0\label{transverseStokesEqn}\\
 -(i\rho \omega + \eta q^2)\tilde{v}_z + \eta \tilde{v_z}_{,zz} - \tilde{p}_{,z} + F_z \delta (z - z_0) &=& 0\label{axialStokesEqn}\\
 iq \tilde{v}_l+ \tilde{v}_{z,z} &=& 0 \label{incompressibilityEqn} 
\end{eqnarray}

The pressure in Eq. \eqref{longitudinalStokesEqn} can be eliminated using Eq. \eqref{axialStokesEqn}. Since the continuity equation \eqref{incompressibilityEqn} gives a direct relation between the components $\tilde{v}_l$ and $\tilde{v_z}$, the following fourth-order differential equation for $v_z$ is obtained
\begin{equation}
\begin{split}
 \tilde{v}_{z,zzzz} - (2q^2+i\lambda^2) \tilde{v}_{z,zz} + q^2(q^2+i\lambda^2) \tilde{v}_z =\\ \frac{q^2}{\eta} F_z \delta (z - z_0) + \frac{iqF_l}{\eta} \delta'(z-z_0),
 \label{differentialEqnVz}
\end{split}
\end{equation}
 where $\delta'$ is the derivative of the delta Dirac function, satisfying the property $x \delta'(x) = -\delta (x)$ for a real $x$, and $\lambda^2 = \rho \omega / \eta$.
 
\subsection{Boundary conditions}\label{boundaryConditionsSection}
 
\subsubsection{Velocity boundary conditions}
 
At the interface $z=0$, the velocity components are continuous 
\begin{equation}
\left[ \tilde{v}_{\alpha} \right] = 0,
\end{equation}
where $\alpha \in \{l,t,z\}$ and $[f] = f(z = 0^{+}) - f(z = 0^{-})$ denotes the jump of a quantity $f$ across the interface. In addition, the no-slip condition Eq.~\eqref{eqn:noSlip} gives
\begin{equation}
 \tilde{v}_{\alpha} (q,z=0, \omega) = i \omega \tilde{u}_{\alpha} (q, \omega).
 \label{eqn:noSlipFourier}
\end{equation}

\subsubsection{Tangential stress jump} 

The presence of the membrane leads to elastic stresses which, in equilibrium, are balanced by a jump in the fluid stress across the membrane:
\begin{equation}
 [{\sigma}_{z \alpha} ] = [\eta(v_{z,\alpha} + v_{\alpha,z}) ] = {\Delta {f}^{\alpha}},
\end{equation}
where $\alpha \in \{x,y\}$.
The tangential traction jump $\Delta f^{\alpha}$  for an elastic membrane experiencing a small deformation is given by Eq.~\eqref{eqn:stressJump}.
We mention that only the resistance to shear and area dilatation is relevant to the first order approximation for the tangential traction jump.

Using the transformations given by 
\eqref{transformation} together with the no-slip condition Eq. \eqref{eqn:noSlipFourier}, we straightforwardly express the first and second derivatives of $u_{x}$ and $u_{y}$ in our new orthogonal basis.
After some algebra, the two tangential conditions are  
\begin{eqnarray}
 [\tilde{v}_{t,z}] &=& -i \alpha_\mathrm{s} q^2 \tilde{v_t}|_{z=0},\label{T1}\\
 ~[iq \tilde{v}_z+\tilde{v}_{l,z}] &=& - 4i\alpha q^2 \tilde{v}_l|_{z=0}\label{T2draft},
\end{eqnarray}
where 
\begin{equation}
\alpha_\mathrm{s} = \kappa_\mathrm{s}/3 \eta \omega
\end{equation}
is a characteristic length for shear and 
\begin{equation}
\alpha=\alpha_\mathrm{s}/B = (\kappa_\mathrm{s}+\kappa_\mathrm{a})/6\eta\omega
\label{defAlpha}
\end{equation}
with $B=2/(1+C)$.

Eq. \eqref{T1} gives the jump condition at the interface for the transverse velocity component $\tilde{v_t}$. Note that the latter is independent of area-dilatation, whereas both $\kappa_\mathrm{s}$ and $\kappa_\mathrm{a}$ are involved in the longitudinal and the normal velocities.
Eq. \eqref{T2draft} can be written by employing the incompressibility equation \eqref{incompressibilityEqn} together with the continuity of the normal velocity across the interface as
\begin{equation}
 [\tilde{v}_{z,zz}] = -4i \alpha q^2 \tilde{v}_{z,z}|_{z=0}\label{T2}.
\end{equation}

\subsubsection{Normal stress jump}

The  normal-normal component of the jump in the stress tensor reads
\begin{equation}
  [ {\sigma}_{zz} ] = [-p + 2\eta v_{z,z} ] =  {\Delta f^{z}}.
\end{equation}

Only the bending effect is present in $\Delta f^{z}$ to the first order, as it can be seen from Eq. \eqref{eqn:stressJump}. Using the incompressibility equation \eqref{incompressibilityEqn} and the continuity of the longitudinal velocity component across the interface, the normal stress jump in Fourier space reads 
\begin{equation}
[ \tilde{v}_{z,zzz} ] = 4 i \alpha_\mathrm{b}^3 q^6 \tilde{v}_z|_{z=0}, \label{N}
\end{equation}
where 
\begin{equation}
 \alpha_\mathrm{b} =   \sqrt[3]{\frac{\kappa_\mathrm{b}}{4 \eta \omega}},
\end{equation}
is a characteristic length for bending. 


\subsection{Green functions}

The Green's functions are tensorial quantities which describe the fluid velocity in direction $\alpha$ 
\begin{equation}
\tilde{v}_{\alpha} = \mathcal{\tilde{G}}_{\alpha \beta} \tilde{F}_{\beta},
\label{greenFctDef}
\end{equation}
for $\alpha, \beta \in \{l,t,z\}$. 
For computing the particle mobilities the relevant quantities are the diagonal components $\mathcal{\tilde{G}}_{tt}$, $\mathcal{\tilde{G}}_{zz}$, and $\mathcal{\tilde{G}}_{ll}$ which can be derived by solving first the independent Eq.~\eqref{transverseStokesEqn} for $\mathcal{\tilde{G}}_{tt}$, then Eq.~\eqref{differentialEqnVz} for $\mathcal{\tilde{G}}_{zz}$ and finally obtaining $\mathcal{\tilde{G}}_{ll}$ from solving Eq. \eqref{differentialEqnVz} and employing the incompressibility condition \eqref{incompressibilityEqn} as detailed in the following.

\subsubsection{Transverse-transverse component}

Let us denote by $K$ the principal square root of $q^2+i\lambda^2$, i.e. 
\begin{equation}
K =\sqrt{\frac{q^2+\sqrt{q^4+\lambda^4}}{2}} + i \sqrt{\frac{-q^2+\sqrt{q^4+\lambda^4}}{2}}.
\end{equation}
Note that for the steady Stokes equations, $\lambda = 0$ and therefore  $K=q$. The general solution of Eq. \eqref{transverseStokesEqn} for the transverse velocity component is
\begin{equation}
\tilde{v}_t = 
 \begin{cases} 
 A  e^{-Kz} & \mbox{for } z > z_0, \label{firstTransverseEqn}\\ 
 B e^{Kz} + C e^{-Kz} & \mbox{for } 0<z<z_0, \\ 
 D e^{Kz} & \mbox{for } z<0.
 \end{cases}
\end{equation}

The integration constants $A$-$D$ are determined by the boundary conditions. $\tilde{v}_t$ is continuous at $z=z_0$, whereas the first derivative is discontinuous due to the delta Dirac function,
\begin{equation}
  \tilde{v}_{t,z} | _{z = z_0^{+}} -  \tilde{v}_{t,z} | _{z = z_0^{-}} = -\frac{F_t}{\eta}.
\end{equation}

In order to evaluate the four constants, two additional equations must be provided. By applying the continuity of the transverse velocity component at the interface together with the tangential traction jump given by Eq. \eqref{T1}, we find that the transverse-transverse component of the Green function is given by
\begin{equation}
 \tilde{\mathcal{G}}_{tt} = \frac{1}{2 \eta K} \left( e^{-K|z-z_0|} + \frac{i\alpha_\mathrm{s} q^2}{2K-i\alpha_\mathrm{s} q^2} e^{-K(z+z_0)}  \right),
 \label{GttPunsteady}
\end{equation}
for $z \ge 0$ and by
\begin{equation}
 \tilde{\mathcal{G}}_{tt} = \frac{1}{\eta} \frac{1}{2K-i \alpha_\mathrm{s} q^2} e^{-K(z_0-z)},
 \label{GttNunsteady}
\end{equation}
for $z \le 0$.
For the steady Stokes equations, the solution reads 
\begin{equation}
 \tilde{\mathcal{G}}_{tt} = \frac{1}{2 \eta q} \left( e^{-q|z-z_0|} + \frac{i\alpha_\mathrm{s} q}{2-i\alpha_\mathrm{s} q} e^{-q(z+z_0)}  \right),
 \label{GttP}
\end{equation}
for $z \ge 0$ and
\begin{equation}
 \tilde{\mathcal{G}}_{tt} = \frac{1}{\eta q} \frac{1}{2-i \alpha_\mathrm{s} q} e^{-q(z_0-z)},
 \label{GttN}
\end{equation}
for $z \le 0$.

\subsubsection{Normal-normal component}
As we are interested here in $ \tilde{\mathcal{G}}_{zz}$ we set $\tilde{F}_l=0$ in Eq. \eqref{differentialEqnVz}. The general solution of this fourth order differential equation is
\begin{equation}
\tilde{v}_z = 
 \begin{cases} 
 A e^{-qz} + B e^{-Kz} & \mbox{for } z > z_0, \\ 
 Ce^{qz} + De^{-qz}+E e^{Kz} + F e^{-Kz} & \mbox{for } 0<z<z_0, \\ 
 G e^{qz} + H e^{Kz} & \mbox{for } z<0.
 \end{cases}
 \label{generalSolutionNormal}
\end{equation}
At the singularity position, i.e. at $z = z_0$, the velocity $\tilde{v}_z$ and its first two derivatives are continuous. However, the delta Dirac function imposes the discontinuity of the third derivative
\begin{equation}
 \tilde{v}_{z,zzz} | _{z = z_0^{+}} -  \tilde{v}_{z,zzz} | _{z = z_0^{-}} = \frac{q^2 F_z}{\eta}.
\end{equation}

At the membrane, $\tilde{v}_z$ and its first derivative are continuous.
However, shear and bending impose a discontinuity in the second and third derivatives respectively (Eqs. \eqref{T2} and \eqref{N}).
The system can readily be solved in order to determine the constants. 
The calculations are  straightforward but lengthy and thus omitted here. We find that the normal-normal component  of the Green function is given in a compact form by 
\begin{equation}
\begin{split}
 \tilde{\mathcal{G}}_{zz} &= \frac{q}{2\mu KZS}\bigg( Ke^{-q|z-z_0|}-qe^{-K|z-z_0|} \\
 &+\sgn{(z)} \frac{2i\alpha q^3 K (P-Q)}{ZPQ(2i\alpha q^2-S)} \left( e^{-q|z|}-e^{-K|z|} \right) \\
 &+\frac{2i\alpha_\mathrm{b}^3 q^5(qP-KQ)}{ZPQ(2i\alpha_\mathrm{b}^3 q^5 - KS)}\left( K e^{-q |z|}-qe^{-K|z|} \right) \bigg).
 \label{Gzzunsteady}
 \end{split}
\end{equation}
Here $P=e^{qz_0}$, $Q=e^{Kz_0}$, $S = K+q$ and $Z=K-q$.

For the steady Stokes equations, i.e. by taking the limits when $K\to q$ and $Q\to P$, one gets
\begin{equation}
\begin{split}
 \tilde{\mathcal{G}}_{zz} &= \frac{1}{4 \eta q} 
 \bigg(
 \left( 1+q|z - z_0| \right) e^{-q|z-z_0|}  \\
 &+ \left( \frac{i\alpha z z_0 q^3}{1-i\alpha q} + \frac{i \alpha_\mathrm{b}^3 q^3 (1+qz_0)(1+qz)}{1-i\alpha_\mathrm{b}^3 q^3} \right) e^{-q(z+z_0)}  
 \bigg),
 \label{GzzP}
\end{split}
\end{equation}
for $z\ge0$ and
\begin{equation}
 \begin{split}
  \tilde{\mathcal{G}}_{zz} &= \frac{1}{4 \eta q} 
 \bigg(
  1+q(z_0 - z)    \\
  &+ \frac{i \alpha z z_0 q^3}{1-i\alpha q} + \frac{i \alpha_\mathrm{b}^3 q^3 (1+q z_0)(1-qz)}{1 - i \alpha_\mathrm{b}^3 q^3}
 \bigg) e^{-q(z_0-z)}, 
\label{GzzN}
 \end{split}
\end{equation}
for $z \le 0$.
Note that both the shear and the bending moduli are involved in the normal-normal component of the Green functions.

\subsubsection{Longitudinal-longitudinal component}

When the normal force $F_z$ is set to zero in  Eq. \eqref{differentialEqnVz}, and only a tangential force $F_l$ is applied, the derivative of the Dirac function imposes the discontinuity of the second derivative at $z=z_0$, whereas the third derivative is continuous.
We have
\begin{equation}
 \tilde{v}_{z,zz} | _{z = z_0^{+}} -  \tilde{v}_{z,zz} | _{z = z_0^{-}} = \frac{iqF_l}{\eta}.
\end{equation}
After solving Eq. \eqref{differentialEqnVz} for the normal velocity $\tilde{v}_z$, the longitudinal velocity $\tilde{v}_l$ can directly be obtained thanks to the incompressibility equation \eqref{incompressibilityEqn}.
We find that the longitudinal-longitudinal component $\tilde{\mathcal{G}}_{ll}$ is 
\begin{equation}
\begin{split}
 \tilde{\mathcal{G}}_{ll} &= \frac{1}{2\eta ZS}\bigg( Ke^{-K|z-z_0|}-qe^{-q|z-z_0|} \\
 &+ \frac{2i\alpha q^2(KP-qQ)}{ZPQ(2i\alpha q^2-S)} \left( qe^{-q|z|}-Ke^{-K|z|} \right) \\
 &+\sgn{(z)} \frac{2i\alpha_\mathrm{b}^3 q^6 K(P-Q)}{ZPQ(2i\alpha_\mathrm{b}^3 q^5 - KS)}\left( e^{-q |z|}-e^{-K|z|} \right) \bigg).
 \label{Gllunsteady}
\end{split}
\end{equation}

When the steady Stokes equations are considered, one simply gets
\begin{equation}
 \begin{split}
  \tilde{\mathcal{G}}_{ll} &= \frac{1}{4 \eta q} 
 \bigg(
(1-q |z - z_0|) e^{-q|z-z_0|}  \\
 &+ \left( \frac{i\alpha q (1-q z_0)(1-qz)}{1-i\alpha q} + \frac{i z z_0 \alpha_\mathrm{b}^3 q^5}{1-i \alpha_\mathrm{b}^3 q^3} \right) e^{-q(z+z_0)}  
 \bigg),\label{GllP}
 \end{split}
\end{equation}
 for $z\ge0$ and
\begin{equation}
\begin{split}
 \tilde{\mathcal{G}}_{ll} &= \frac{1}{4 \eta q} 
 \bigg(
  1 - q(z_0 - z)  \\
  &+ \frac{i\alpha q (1+qz)(1-qz_0)}{1-i\alpha q} + \frac{i z z_0 \alpha_\mathrm{b}^3 q^5}{1-i\alpha_\mathrm{b}^3 q^3}
 \bigg)e^{-q(z_0-z)} ,\label{GllN}
\end{split}
\end{equation}
for $z \le 0$.


\subsection{Particle mobilities}

We now obtain the mobility corrections defined in Eq.~\eqref{eqn:defMobility} and given specifically in Eqs.~\eqref{eqn:mu_para_s_unsteady}-\eqref{eqn:mu_perp_b_unsteady} (including the inertial term) and Eqs.~\eqref{eqn:paraShear}-\eqref{eqn:perpBen} (without fluid inertia) of the main text. 
For this, using Eq.~\eqref{transformation} on Eq.~\eqref{greenFctDef}, one derives the transformation of the tensorial Green's functions back to Cartesian directions:
\begin{eqnarray}
\tilde{\mathcal{G}}_{xx} (\vect{q},z,\omega) &=& \frac{q_y^2}{q^2}\tilde{\mathcal{G}}_{tt} (\vect{q},z,\omega) + \frac{q_x^2}{q^2}\tilde{\mathcal{G}}_{ll} (\vect{q},z,\omega),\label{G_xx_backToCartesian}\\
\tilde{\mathcal{G}}_{yy} (\vect{q},z,\omega) &=& \frac{q_x^2}{q^2}\tilde{\mathcal{G}}_{tt} (\vect{q},z,\omega) + \frac{q_y^2}{q^2}\tilde{\mathcal{G}}_{ll} (\vect{q},z,\omega).\label{G_yy_backToCartesian}
\end{eqnarray}

We then subtract the infinite space Green's functions in the Fourier domain which can be obtained via the above derivation with the membrane moduli set to zero, i.e.
\begin{equation}
 \Delta \tilde{\mathcal{G}}_{\gamma \gamma}^{(0)}(\vect{q},z,\omega) = \tilde{\mathcal{G}}_{\gamma \gamma}(\vect{q},z,\omega)- \tilde{\mathcal{G}}_{\gamma \gamma}(\vect{q},z,\omega)|_{\alpha,\alpha_\mathrm{b} = 0},
 \label{mobilityCorrectionDef}
\end{equation}
where $\gamma \in \{x,y,z\}$.
This defines the wave-vector dependent corrections
\begin{eqnarray}
\Delta \tilde{\mathcal{G}}_{\parallel}\left(\vect{q},z,\omega\right) &=&  \tilde{\mathcal{G}}_{xx}\left(\vect{q},z,\omega\right) - \tilde{\mathcal{G}}_{xx}^{(0)}\left(\vect{q},z,\omega\right),\notag\\
 &=&  \tilde{\mathcal{G}}_{yy}\left(\vect{q},z,\omega\right) - \tilde{\mathcal{G}}_{yy}^{(0)}\left(\vect{q},z,\omega\right),\notag\\
\Delta \tilde{\mathcal{G}}_{\perp}\left({q},z,\omega\right) &=&  \tilde{\mathcal{G}}_{zz}\left({q},z,\omega\right) - \tilde{\mathcal{G}}_{zz}^{(0)}\left({q},z,\omega\right).
\label{eqn:deltaGofq}
\end{eqnarray}

Due to the point-particle approximation it is sufficient to obtain the fluid velocity at the particle position which is equal to the velocity of the particle itself. Instead of the full inverse Fourier transform of the Green's functions to real space coordinates ($\vect{\rho}$, $z$), we can thus limit ourselves to evaluate the inverse Fourier transform of Eqs.~\eqref{eqn:deltaGofq} at ($\vect{\rho}=0$, $z=z_0$). By passage to polar coordinates $q_x = q\cos \phi$ and $q_y = q \sin \phi$, the correction in the particle  mobility to the first order of $R/z_0$ can be obtained
\begin{eqnarray}
 \Delta {\mu}_\parallel(\omega) &=& \frac{1}{(2\pi)^2}
 \int_{0}^{2\pi} \int_{0}^{\infty} \Delta \tilde{\mathcal{G}}_{\parallel} (q,\phi, z=z_0, \omega) q d  q d  \phi\notag\\
  \Delta {\mu}_\perp(\omega) &=& \frac{1}{2\pi}
  \int_{0}^{\infty} \Delta \tilde{\mathcal{G}}_{\perp} (q, z=z_0, \omega) q d q,\notag\\
 \label{mobilityCorrectionFormulaPointParticle}
\end{eqnarray}
which directly lead to Eqs.~\eqref{eqn:mu_para_s_unsteady}-\eqref{eqn:mu_perp_b_unsteady} of the main text. A similar procedure can be followed for the steady case where the fluid inertia is neglected leading to Eqs.~\eqref{eqn:paraShear}-\eqref{eqn:perpBen}.

\section{COMPUTING MEAN-SQUARE-DISPLACEMENTS FROM PARTICLE MOBILITIES}
\label{sec:diffusion}

\subsection{Time dependent mobility corrections}

A crucial step in order to compute the mean-square-displacements as described in the following section is to transform the frequency-dependent particle mobilities back to the time domain. As shown in the Supporting Information, the inertial contribution to the mobility correction is negligible for realistic scenarios and we therefore restrict ourselves from now on to the case $\sigma = 0$.
For the sake of simplicity, we do not start from the real-space particle mobilities given in Eqs.~\eqref{eqn:paraShear}-\eqref{eqn:perpBen}, but instead depart from the wave-vector-dependent Green's functions in Eq.~\eqref{eqn:deltaGofq} to perform first an inverse Fourier transform in time followed by an inverse Fourier transform in space. 
Note that the inverse order is possible for the shear-related part, but the calculations are much more complicated.

\subsubsection{Parallel mobility}

\textit{Shear effect}. Considering only the part due to shear resistance in Eqs.~\eqref{GllP} and \eqref{GttP} and using Eq. \eqref{eqn:deltaGofq} with \eqref{mobilityCorrectionDef},
we find after passing to polar coordinates:
\begin{equation}
\begin{split}
 \Delta \tilde{\mathcal{G}}_{\parallel, \mathrm{s}}(q,\phi,\omega)|_{z=z_0} &= \frac{i z_0 e^{-2q z_0}}{2\eta} \bigg( \frac{\sin^2\phi}{T_\mathrm{s}\omega -iq z_0} \\
 &+ \frac{(1-q z_0)^2 \cos^2 \phi}{B T_\mathrm{s} \omega-2iqz_0} \bigg),
\end{split}
\end{equation}
where $T_\mathrm{s}=6z_0 \eta / \kappa_\mathrm{s}$ is a characteristic time for shear. The temporal inverse Fourier transform reads
\begin{equation}
\begin{split}
 \Delta \tilde{\mathcal{G}}_{\parallel, \mathrm{s}}(q,\phi,t)|_{z=z_0} &= -\frac{z_0 e^{-2q z_0} \theta(t)}{2\eta T_\mathrm{s}} \bigg( e^{-\frac{qz_0 t}{T_\mathrm{s}}}\sin^2 \phi \\
 &+ \frac{(1-qz_0)^2}{B} e^{\frac{-2qz_0t}{B T_\mathrm{s}}}\cos^2\phi \bigg).
\end{split}
\end{equation}

An exact expression of the time dependent mobility correction due to shear in the parallel case can then be obtained by spatial inverse Fourier transform 
\begin{equation}
 \frac{\Delta \mu_{\parallel, \mathrm{s}}({\tau})}{\mu_0} = -\frac{3}{32}\frac{R}{z_0} \frac{\theta(\tau)}{T_\mathrm{s}}
 \frac{N_B(\tau)}{ (2+{\tau})^2 ({\tau}+B)^4}
  \label{DeltaMuTimeParaShear}
\end{equation}
where $\tau = t/T_\mathrm{s}$, and again $B=2/(1+C)$. $\theta (t)$ denotes the Heaviside step function, with $\theta(0)=1/2$ and 
\begin{equation}
\begin{split}
 N_B(\tau) &= 4B^3(1+2B)+36B^3{\tau} +B(B^2+48B+8){\tau}^2\\
 &+ 40B{\tau}^3+2(B+4){\tau}^4.
 \end{split}
\end{equation}

\textit{Bending effect}. Considering the part due to bending resistance we obtain
\begin{equation}
 \Delta \tilde{\mathcal{G}}_{\parallel, \mathrm{b}} (q,\phi,\omega)|_{z=z_0} = \frac{\cos^2 \phi}{4\eta} \frac{i q^4 z_0^5}{T_\mathrm{b} \omega - i q^3 z_0^3} e^{-2q z_0},
\end{equation}
to give after applying the temporal inverse Fourier transform
\begin{equation}
 \Delta \tilde{\mathcal{G}}_{\parallel,\mathrm{b}} (q,\phi,t)|_{z=z_0} = -\frac{q^4 z_0^5 \theta (t) \cos^2 \phi }{4\eta T_\mathrm{b}} e^{-2q z_0 - \frac{t q^3 z_0^3}{T_\mathrm{b}}}.
\end{equation}

The time dependent mobility can immediately be obtained after applying the inverse Fourier transform
\begin{equation}
 \frac{\Delta \mu_{\parallel,\mathrm{b}} (t)}{\mu_0} = - \frac{3}{8} \frac{a}{z_0} \frac{\theta (t)}{T_\mathrm{b}} \int_{0}^{\infty} u^5 e^{-2u-\frac{t}{T_\mathrm{b}} u^3} d u, 
 \label{bendingMobilityCorrectionTimePara}
\end{equation}
where $T_\mathrm{b} = 4\eta z_0^3 / \kappa_\mathrm{b}$. The presence of $u^3$ in the exponential argument makes the analytical evaluation of this integral impossible. To overcome this difficulty, we evaluate the integral numerically and fit the result (as a function of $t$) with an analytical empirical form which is necessary to proceed further. This procedure is known as the Batchelor parametrization \cite{batchelor50}. It can be shown that the integral decays following a $t^{-2}$ law for larger times. Therefore, we can write
\begin{equation}
 \frac{ \Delta \mu_{\parallel,\mathrm{b}} (\tau_{\parallel,\mathrm{b}})}{\mu_0} = -\frac{45}{64} \frac{R}{z_0} \frac{\theta(\tau_{\parallel,\mathrm{b}})}{T_\mathrm{b}} \frac{1}{\left( \tau_{\parallel,\mathrm{b}}^{p}+1 \right)^{\frac{2}{p}}},
 \label{DeltaMuTimeParaBending}
\end{equation}
where $p=1/2$ is the fitting parameter and $\tau_{\parallel,\mathrm{b}} = (5/2)(t/T_\mathrm{b})$. A comparison between the numerically obtained value of the integral and the fitting formula is presented in the SI, where a good agreement is obtained.


\subsubsection{Perpendicular motion}

\textit{Shear effect.} 
Considering only the part due to shear resistance in Eq.~\eqref{GzzP} and using Eq. \eqref{eqn:deltaGofq} with Eq.~\eqref{mobilityCorrectionDef} we find after passing to polar coordinates:
\begin{equation}
 \Delta \tilde{\mathcal{G}}_{\perp, \mathrm{s}} (q,\omega)|_{z=z_0} = \frac{i q^2 z_0^3}{2\eta} \frac{e^{-2q z_0}}{B T_\mathrm{s} \omega - 2iq z_0}.
\end{equation}

The computation of the temporal inverse Fourier transform leads to
\begin{equation}
  \Delta \tilde{\mathcal{G}}_{\perp, \mathrm{s}} (q,t)|_{z=z_0} = -\frac{q^2 z_0^3 \theta(t)}{2\eta B T_\mathrm{s}} e^{-2q z_0 \left( 1+\frac{t}{B T_\mathrm{s}} \right)}.
\end{equation}

After applying the spatial inverse Fourier transform to this equation, we find that the time dependent mobility correction due to shear reads
\begin{equation}
 \frac{\Delta \mu_{\perp, \mathrm{s}}(\tau)}{\mu_0} = -\frac{9}{16} \frac{R}{z_0} \frac{\theta(\tau)}{T_\mathrm{s}} \frac{B^3}{\left(\tau+B\right)^4}.
 \label{DeltaMuTimePerpShear}
\end{equation}

\textit{Bending effect}. Considering only the part due to bending resistance we obtain
\begin{equation}
 \Delta \tilde{\mathcal{G}}_{\perp, \mathrm{b}} (q,\omega)|_{z=z_0} = \frac{i q^2 z_0^3 (1+q z_0)^2}{4\eta} \frac{e^{-2q z_0}}{T_\mathrm{b} \omega - i q^3 z_0 ^3}.
\end{equation}
The temporal inverse Fourier transform is
\begin{equation}
  \Delta \tilde{\mathcal{G}}_{\perp, \mathrm{b}} (q,t)|_{z=z_0} = -\frac{q^2 z_0^3 (1+q z_0)^2 \theta(t)}{4\eta T_\mathrm{b}} e^{-2q z_0 - \frac{t q^3 z_0^3}{T_\mathrm{b}}}.
\end{equation}

After Fourier-transform in space, the time dependent mobility correction due to bending is expressed by the following improper integral
\begin{equation}
 \frac{\Delta {\mu}_{\perp,\mathrm{b}}(t)}{\mu_0} = -\frac{3}{4} \frac{a}{z_0} \frac{\theta(t)}{T_\mathrm{b}} \int_0^{\infty} u^3(1+u)^2 e^{-2u- \frac{t u^3}{T_\mathrm{b}} } d  u.
 \label{bendingMobilityCorrectionTimePerp}
\end{equation}
As above, we use the Batchelor parametrization \cite{batchelor50} to represent the integral. At $t=0$, the integral above can be solved analytically, and it is equal to $15/4$. At larger times, the integral decays monotonically following a $t^{-4/3}$ law.
We set
\begin{equation}
  \frac{\Delta {\mu}_{\perp, \mathrm{b}}(\tau_{\perp,\mathrm{b}})}{\mu_0} = - \frac{45}{16} \frac{R}{z_0} \frac{\theta(\tau_{\mathrm{b}})}{T_\mathrm{b}} \frac{1}{\left( \tau_{\perp,\mathrm{b}}^{p} + 1 \right)^{\frac{4}{3p}}},
   \label{DeltaMuTimePerpBending}
\end{equation}
where $\tau_{\perp,\mathrm{b}} = (9 \pi/4) (t/T_\mathrm{b})$ and $p = 2/3$ is a fitting parameter, governing the evolution of the mobility correction at short times.
Again, the fitting formula and the numerical solution are in excellent agreement as seen in the Supporting Information.


\subsection{Mean-square-displacements}

The dynamics of a Brownian particle are governed by the generalized Langevin equation \cite{kubo66} 
\begin{equation}
m \frac{d  {v}_{\alpha}}{d  t} = -\int_{-\infty}^{t} \gamma_{\alpha}(t-t'){v}_{\alpha}(t') d  t' + {F} (t),
 \label{generalizedLangevinEqn}
\end{equation}
where $m$ is the particle mass and ${v}_{\alpha}$ is its velocity in direction $\alpha = \parallel, \perp$. 
$\gamma_{\alpha}(t)$ denotes the time dependent friction retardation function (expressed in $\operatorname{kg} / \operatorname{s}^{2}$), and ${F}$ is the random force which is zero on average. The random force results from the impacts with the fluid molecules due to the thermal fluctuation. The relation between the mobility and the friction function is given by \cite[Eq. (1.6.4) p. 32]{kubo85}
\footnote{Note that the retardation function as defined by Kubo in \cite{kubo85} does not incorporate the particle mass $m$.
I.e. $\gamma(t)$ as it appears in the generalized Langevin equation is expressed in s$^{-2}$ while ours in kg/s$^2$.
That is the reason why $m$ appears as a factor in Eq. (1.6.4) p. 32 and Eq. (1.6.14) p. 34.}
\begin{equation}
 \mu_{\alpha} (\omega) = \frac{1}{im\omega+\gamma_{\alpha}[\omega]}.\label{muDef}
\end{equation}
where
 $\gamma_{\alpha}[\omega]$ is the one-sided Fourier transform of the retardation function defined by
\begin{equation}
 \gamma_{\alpha}[\omega] = \int_{0}^{\infty} \gamma_{\alpha}(t) e^{-i \omega t} d  t.
\end{equation}


The frictional forces and the random forces are not independent quantities, but are related to each other via the fluctuation-dissipation theorem (FDT) \cite{kubo66}. According to the FDT, the velocity autocorrelation function (VACF) has the following expression \cite[Eq. (1.6.14) p. 34]{kubo85}
\begin{equation}
 \phi_{v,\alpha} (t) \coloneqq \langle v_\alpha(0)v_\alpha(t) \rangle = \frac{k_{\text{B}} T}{2\pi} \int_{-\infty}^{\infty} 
 \mu_\alpha (\omega) e^{i\omega t} d  \omega.
 \label{ACFInt}
\end{equation}


In the overdamped regime, i.e. for a massless particle, Eq. \eqref{ACFInt} is reduced to
\begin{equation}
 \phi_{v,\alpha} (t) = D_0 \left( 2\delta(t) + \frac{\Delta \mu_\alpha(t)}{\mu_0} \right),
 \label{ACFDef}
\end{equation}
where  $D_0  = k_\text{B} T \mu_0$, is the bulk diffusion coefficient given by the Einstein relation \cite{einstein05}. 

Next, the particle MSDs can be computed knowing the VACF as \cite{kubo66}
\begin{eqnarray}
 \langle x(t)^2 \rangle &=& 2\int_{0}^{t} (t-s)\phi_{v,\parallel} (s) d s\notag\\
 \langle z(t)^2 \rangle &=& 2\int_{0}^{t} (t-s)\phi_{v,\perp} (s) d s,
 \label{MSDDef}
\end{eqnarray}
which can be conveniently split up into a bulk contribution and a correction defined by:
\begin{align}
 \Delta_{\parallel} (t) &= 1-\frac{\langle x(t)^2 \rangle}{2 D_0 t} \, , \\
\Delta_{\perp} (t) &= 1-\frac{\langle z(t)^2 \rangle}{2 D_0 t}.
\end{align}

By inserting the time-dependent mobility corrections derived in Eqs.~\eqref{DeltaMuTimeParaShear}, \eqref{DeltaMuTimeParaBending}, \eqref{DeltaMuTimePerpShear}, \eqref{DeltaMuTimePerpBending} in Eq.~(\ref{ACFDef}) and using Eqs.~(\ref{MSDDef}) we obtain analytical expressions for the excess mean-square-displacement as follows:
\begin{widetext}
\begin{eqnarray}
 \Delta_{\perp, \mathrm{s}} (\tau) &=& \frac{3}{16} \frac{R}{z_0} \frac{\tau (3B+2\tau)}{2(B+\tau)^2},\label{DeltaDiffusionPerpShear}\\
 \Delta_{\perp, \mathrm{b}}(\tau_{\perp, \mathrm{b}}) &=& \frac{15}{16}\frac{R}{z_0} \frac{2}{\pi} \left(\arctan \tau_{\perp, \mathrm{b}}^{\frac{1}{3}}- \frac{2}{\tau_{\perp, \mathrm{b}}^{\frac{1}{3}}} + \frac{2}{\tau_{\perp, \mathrm{b}}} \ln \left(1+\tau_{\perp, \mathrm{b}}^{\frac{2}{3}} \right) \right),\label{DeltaDiffusionPerpBending}\\
 \Delta_{\parallel, \mathrm{s}}(\tau) &=& \frac{15}{32} \frac{R}{z_0} \frac{1}{10} 
 \left( \frac{(2\tau+3B)(5\tau+4B)}{(B+\tau)^2} - \frac{4B}{\tau} \ln\left( 1+\frac{\tau}{B}\right) - \frac{16}{\tau} \ln \left( 1+\frac{\tau}{2}\right) \right),
 \label{excessPara}\\
 \Delta_{\parallel, \mathrm{b}}(\tau_{\parallel, \mathrm{b}}) &=& \frac{3}{32}\frac{R}{z_0} \left( \frac{\tau_{\parallel, \mathrm{b}}^{3/2}+2\tau_{\parallel, \mathrm{b}}+9\sqrt{\tau_{\parallel, \mathrm{b}}}+6}{\sqrt{\tau_{\parallel, \mathrm{b}}} (1+\sqrt{\tau_{\parallel, \mathrm{b}}})^2} - \frac{6}{\tau_{\parallel, \mathrm{b}}}\ln \left(1+\sqrt{\tau_{\parallel, \mathrm{b}}} \right)\right).
 \label{DeltaDiffusionParaBending}
\end{eqnarray}
\end{widetext}


\input{All.bbl}

\widetext

\clearpage

\widetext

\begin{center}
\textbf{\large Supplemental Materials for: Elastic cell membranes induce long-lived anomalous thermal diffusion on nearby particles}

\vspace{1cm}

{Abdallah Daddi-Moussa-Ider, Achim Guckenberger and Stephan Gekle}

\vspace{0.2cm}
{Biofluid Simulation and Modeling, Fachbereich Physik, Universit\"at Bayreuth}
\end{center}

\vspace{1cm}

\setcounter{equation}{0}
\setcounter{figure}{0}
\setcounter{table}{0}
\setcounter{page}{1}
\makeatletter
\renewcommand{\theequation}{S\arabic{equation}}
\renewcommand{\thefigure}{S\arabic{figure}}
\renewcommand{\bibnumfmt}[1]{[S#1]}
\renewcommand{\citenumfont}[1]{S#1}

\end{appendix}

\setcounter{section}{0}

\input{SIsuite}

\end{document}

%% file: All.bbl
%

%% file: SIsuite.tex
\section{Simulation methods}

\subsection{Boundary-Integral simulations}

In the creeping flow approximation, i.e. in the low Reynolds number regime, the fluid motion is governed by the steady Stokes equations. The equations are formulated as integral equations \cite{Pozrikidis_book} which are solved numerically after generating a triangulated mesh on the boundaries. The goal is to determine the particle translational and rotational velocities when a given force and torque are applied on its surface.
This setup is commonly referred to as solving for the mobility problem \cite{kohr_2004}. 

The moving particle disturbs the fluid velocity field in its vicinity. As a result, the membrane is deformed and exerts a force on the surrounding fluid in order to regain its equilibrium configuration. To solve for both the particle and the membrane velocities, we use a completed double layer boundary integral equation method (CDLBIEM) \cite{Zhao_2012} which is able to treat a perfectly rigid extended solid particle
\begin{eqnarray}
 v_j (\vect{x}) &=& H_j (\vect{x}), ~~\vect{x} \in S_\mathrm{m} \label{BIM1} \\
 \frac{1}{2} \phi_j (\vect{x}) + \sum_{i = 1}^{6} \varphi_j^{(i)} \langle \boldsymbol{\varphi}^{(i)}, \boldsymbol{\phi} \rangle (\vect{x}) &=& H_j (\vect{x}), ~~\vect{x} \in S_\mathrm{p}\label{BIM2}
\end{eqnarray}
where $S_{\mathrm{m}}$ and $S_{\mathrm{p}}$ denote the membrane and the particle surface, respectively, $\vect{v}$ is the membrane velocity and $\boldsymbol{\phi} (\vect{x})$ is the double layer density function. The $\boldsymbol{\varphi}^{(i)} (\vect{x})$ are known functions that depend on the particle position and geometry \cite{Kim_2005_book}. Finally, the function $H_j$ is
\begin{equation}
 H_j (\vect{x}) = -\frac{1}{8 \pi \mu} (N_\mathrm{m} \Delta f)_j (\vect{x})
		      -\frac{1}{8 \pi} (K_\mathrm{p} \phi)_j (\vect{x})\\
		      +\frac{1}{8 \pi \mu} (G_{jk} F_k + R_{jk} M_k) (\vect{x}). 
\end{equation}

The operator $N_{\mathrm{m}}$ denotes the single layer integral over $S_{\mathrm{m}}$ while $K_{\mathrm{p}}$ is the double layer integral over $S_{\mathrm{p}}$, respectively defined by
\begin{eqnarray}
 (N_\mathrm{m} \Delta f)_j (\vect{x}) &=& \int_{S_\mathrm{m}} \Delta f_i(\vect{y}) G_{ij}(\vect{y},\vect{x}) d  S(\vect{y}),\\
 (K_\mathrm{p} \phi)_j (\vect{x}) &=& \oint_{S_\mathrm{p}} \phi_i(\vect{y}) T_{ijk}(\vect{y},\vect{x}) \vect{n}_k (\vect{y}) d  S(\vect{y}),
\end{eqnarray}
where $\vect{y}\in S_{\mathrm{m}} \cup S_{\mathrm{p}}$ and $\vect{n}$ is the normal vector to the surface.
The traction jump across the membrane is $\Delta \vect{f}$, which is determined from the membrane energetics (see below). 
$\vect{G}$, $\vect{T}$ and $\vect{R}$ are the stokeslet, the stresslet and the rotlet respectively (known tensors), and $\vect{F}$ and $\vect{M}$ are the force and the torque exerted on the particle. After discretization, equations \eqref{BIM1} and \eqref{BIM2} form a linear system that is solved with GMRES \cite{Saad_1986}. The particle translation and rotation velocities can directly be computed from the double layer density $\boldsymbol{\phi}$ using Eq.~\eqref{BIM2}.

For the mobility simulations depicted in figures~\eqref{mobility_artificial} and \eqref{mobility_diff} of the main text, a spherical particle with radius $R=1$ for (a) and (b) and $R=6.5$ for (c) and (d) is placed a distance 10 above the membrane at $z=0$ (all numbers in this paragraph are given in simulation units). The oscillating force acting on the particle has an amplitude of 10$^{-4}$ and we have checked that doubling the force still leads to the same mobilities (linear response). The membrane is quadratic with a size of $300 \times 300$ and is meshed with 1740 triangles. The mesh has been created with gmsh \cite{Geuzaine_2009} and the triangle size increases towards the outer regions to guarantee a high resolution in the center close to the particle at affordable computational cost. The triangle vertices located on the outer edge of the membrane are constrained with harmonic springs. The spring constant $k$ is chosen equal to the elastic modulus of the membrane $\kappa_{\mathrm{s}}$ in order to mimick the inifinite system considered in the theory. We have checked that variation of $k$ within reasonable bounds does not strongly influence the results.

\subsection{Membrane energetics}\label{membraneMechanics}

\subsubsection{Elastic model}

We use for the membrane the Skalak constitutive law \cite{Skalak_1973} whose areal strain energy density reads \cite{Lac_2004}
\begin{equation}
 w_{\mathrm{s}} = \frac{\kappa_{\mathrm{s}}}{12} (I_1^2 + 2I_1 -2 I_2 + C I_2^2),
\end{equation}
where $\kappa_{\mathrm{s}}$ is the membrane elastic shear modulus,
and $C = \kappa_{\mathrm{a}} / \kappa_{\mathrm{s}}$ is the ratio between the area dilatation and shear moduli. The strain tensor invariants are $I_1$ and $I_2$, with $I_1 = \lambda_1^2+\lambda_2^2-2$ and $I_2 = \lambda_1^2 \lambda_2^2 - 1$. Here $\lambda_1$ and $\lambda_2$ denote the local in-plane principal strains. By integrating the areal energy density $w_{\mathrm{s}}$ over the surface of reference $S_0$, the total strain energy can be computed \cite{Le_2010},
\begin{equation}
 W_{\mathrm{s}} = \int_{S_0} w_{\mathrm{s}} (\vect{x}) d  S (\vect{x}).
\end{equation}

The membrane is  discretized numerically into flat triangles, which are assumed to remain plane even after deformation. We use the finite element approach introduced by Charrier \textit{et al.} \cite{Charrier_1989} in order to compute the membrane force on each discrete node. The relative displacement can then be determined by transforming the deformed and undeformed elements to the same plane \cite{Kruger_2011_paper}. The local principal in-plane ratios $\lambda_1$ and $\lambda_2$ can be computed from the displacement tensor, and consequently the strain energy can be evaluated. 
The elastic force applied  on the flowing fluid by the membrane node $\vect{x}_i$ can be obtained from the virtual work principal,
\begin{equation}
 \vect{F}(\vect{x}_i) = -\frac{\partial  W_{\mathrm{s}}}{\partial \vect{x}_i}.
\end{equation}
To evaluate the traction jump $\Delta \vect{f}$, the force is divided by the area associated with the node $\vect{x}_i$ \cite{Spann_2014}.

\subsubsection{Bending model}

We use the bending energy as given by Helfrich \cite{Helfrich_1973}
\begin{equation}
 W_{\mathrm{b}} =  \frac{\kappa_{\mathrm{b}}}{2} \int_S \left( 2 H(\vect{x}) - 2H_0 (\vect{x}) \right)^2 d  S(\vect{x}),
 \label{helfrichEnergy}
\end{equation}
where $\kappa_\mathrm{b}$ is the bending modulus and $H$ is the mean curvature.
$H_0$ is the mean curvature of the surface of reference.
The traction jump is directly calculated  by evaluating the functional derivative of the bending energy \cite{Pozrikidis_2001_stiff, Laadhari_2010},
\begin{equation}
 \Delta \vect{f} (\vect{x}) =-\kappa_{\mathrm{b}} \left( ( 2H^2-2K+2H_0 H + \Delta )(2H-2H_0) \right) \vect{n},
\end{equation}
where $\Delta$ is the Laplace-Beltrami operator, and $K$ is the Gaussian curvature. The general approach of the numerical discretization of these operators can be found in Meyer \textit{et al.} \cite{Meyer_2003}.


\clearpage

\subsection{Obtaining particle mobilities from numerical simulations}

\begin{figure}[h!]
\includegraphics[width=0.5\columnwidth]{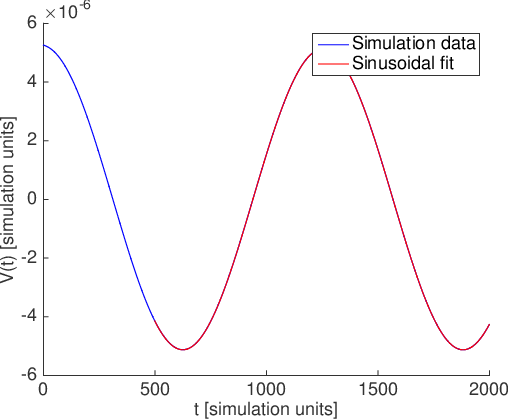}
\caption{The particle velocity as a function of time obtained from the BIM simulations. After an initial transient the velocity oscillates with the same frequency $\omega$ as the driving force: for $t>500$ simulation and fit overlap perfectly.
\label{fig:mobility_fit} }
\end{figure}

To obtain the particle mobilities $\mu(\omega)$ shown in figures~\ref{mobility_artificial} and \ref{mobility_diff} of the main text from boundary integral simulations, we apply a sinusoidal force $F(t)=F_0\cos(\omega t)$ in $x$-direction (for $\Delta\mu_\parallel$) or in $z$-direction (for $\Delta\mu_\perp$). We then record the particle velocity which -- after a short transient -- oscillates with the same frequency as the force as illustrated in figure~\ref{fig:mobility_fit} for $R/z_0=1/10$ and parallel motion. Accordingly, the velocity is fitted with a function $V_0 \cos(\omega t + \delta)$. From the phase shift $\delta$ and the ratio of the amplitudes we then calculate the mobility as
\begin{equation}
\mu_\parallel(\omega) = \frac{V_0}{F_0}e^{i\delta}.
\end{equation}

\clearpage


\section{Effect of the unsteady term}

\subsection{Particle mobilities}

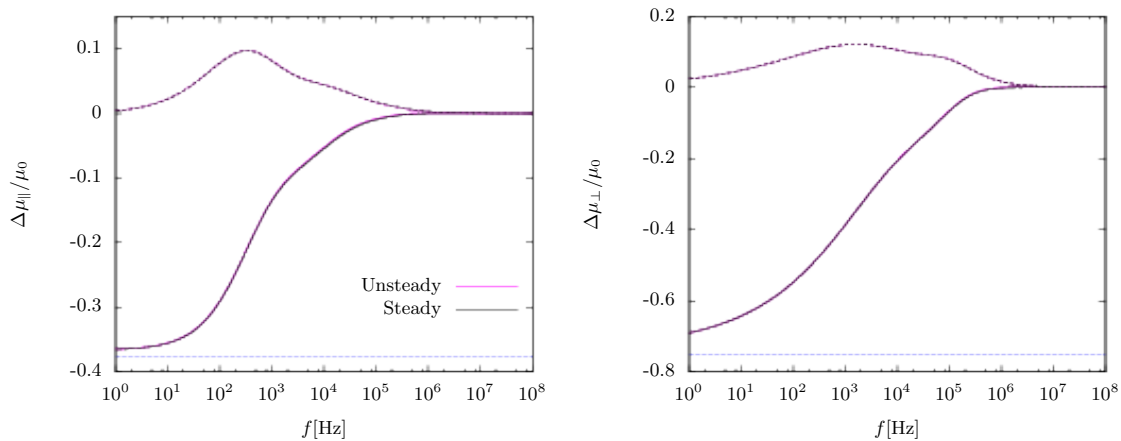
\begin{figure}[h]
  \begin{center}
     \scalebox{.8}{\input{steadyUnsteady}}
     \caption{Comparison between the steady and the unsteady parallel (left) and perpendicular (right) mobility corrections for the physical parameters corresponding to Fig. \ref{mobility_diff} in the main text. The horizontal solid lines are the hard wall limits.
     \label{steadyUnsteady}}
  \end{center}
\end{figure}

In order to investigate the effect of the unsteady term in the Stokes equations, we solve numerically the integrals appearing in the mobility corrections Eqs.~\eqref{eqn:mu_para_s_unsteady}-\eqref{eqn:mu_perp_b_unsteady}. Here we take the same physical parameters as in Fig. \ref{mobility_diff} of the main text and the fluid density $\rho=10^3$ kg/m$^3$. In figure~\ref{steadyUnsteady} we show a comparison between the unsteady mobility corrections and the analytical solutions for the steady case in Eqs.~\eqref{eqn:paraShear}-\eqref{eqn:perpBen} of the main text. We find that the two mobility corrections are almost indistinguishable for the whole range of frequencies.

\subsection{Mean-square displacements}

\begin{figure}[h]
  \begin{center}
     \includegraphics[width=0.5\textwidth]{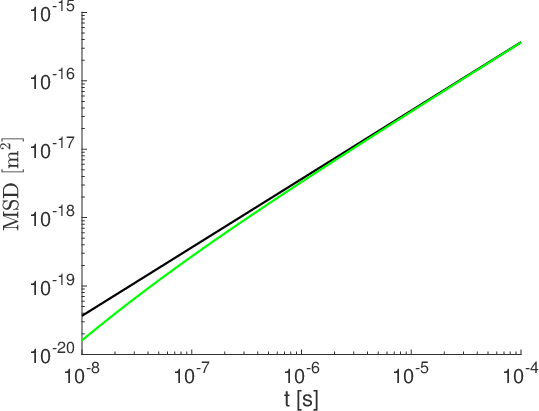}
     \caption{Comparison of the bulk mean-square displacement with (green line) and without (black line) the inertial contribution. Inertial contributions become significant only for times shorter than 1$\mu$s which is smaller than the time range relevant in this work.}
     \label{fig:MSD_unsteady}
  \end{center}
\end{figure}

Having shown in the previous section that the effect of the unsteady inertial term on the mobility correction is negligible, we proceed to consider the influence of the unsteady bulk mobility $\mu_0^u(\omega)$ on the mean-square-displacement. Due to the linearity of the equations it is sufficient to compare the steady-state bulk MSD with the unsteady bulk MSD (the membrane contribution is simply added on top). These can be obtained from Eqs.~\eqref{eqn:MSD_from_mu} of the main text using $\mu(\omega)=\mu_0$ or $\mu(\omega)=\mu_0+\mu_0^u(\omega)$, respectively. The result is shown in figure~\ref{fig:MSD_unsteady}. The influence of the unsteady term is restricted to $t<1\mu$s which is much shorter than the regime where the membrane influences the MSD as can be seen by comparing with figure~\ref{fig:MSD} of the main text.

\section{Relation to previous works}

\subsection{Comparison to the work of Felderhof \cite{felderhof06}}

Here we make contact with the mobility correction due to the elastic in-plane deformations of the membrane as reported by Felderhof \cite{felderhof06} in the case where the two fluids have the same dynamic viscosity, i.e. $\eta=\eta_1 = \eta_2$. The tensor $F(\vect{r}_0, \omega)$ that appears in Eq.~(2.16) of  \cite{felderhof06} is the first order correction in the mobility tensor that is also the subject of our work detailed in the main text. Based on Eqs. (3.5) and (3.6) of \cite{felderhof06} we define (note the $q$-dependence)
\begin{eqnarray}
F_{xx} (q, \omega,h) &=& \frac{N_{xs} (q,\omega,h)}{Y(q,\omega)} + \frac{N_{xp} (q,\omega,h)}{Z(q,\omega)},\\
 F_{zz} (q, \omega,h) &=& \frac{N_{zp} (q,\omega,h)}{Z(q,\omega)}.
\end{eqnarray}
By taking $q_1 = q_2 = \sqrt{q^2+\delta^2}$, where $\delta^2 = -i\omega \rho / \eta$, and carefully taking the limits when $\delta\to 0$, we find after simplifications 
\begin{eqnarray}
 F_{xx} (q, \omega,h) &=& -\frac{1}{8\pi\eta} \frac{i\alpha_\mathrm{s}}{2+i\alpha_{\mathrm{s}} q }e^{-2q h} -\frac{1}{16\pi\eta} \frac{i\alpha_\mathrm{k} (qh - 1)^2}{1 + i\alpha_\mathrm{k} q} e^{-2q h},\label{FF1}\\
 F_{zz} (q, \omega,h) &=& -\frac{1}{8\pi\eta} \frac{i\alpha_\mathrm{k} (qh)^2 }{1+i\alpha_\mathrm{k}q}e^{-2q h},\label{FF2}
\end{eqnarray}
where 
\begin{equation}
 \alpha_\mathrm{k} = \frac{\Lambda}{4\eta \omega} \;\;\;\mathrm{and}\;\;\;  \alpha_\mathrm{s} = \frac{G_\mathrm{s}}{\eta \omega},
\end{equation}
with $\Lambda=E_\mathrm{s}+G_\mathrm{s}$ and the dilatation and shear moduli $E_\mathrm{s}$ and $G_\mathrm{s}$, respectively, as defined in \cite{felderhof06}.

The results obtained in the present work can be cast into a similar form considering that equations (\ref{FF1}) and (\ref{FF2}) correspond to the corrections in the parallel and perpendicular mobilities.
This can be obtained by considering the shear-related parts of Eqs.~\eqref{GttP}, \eqref{GllP} and \eqref{GzzP}, together with Eq. \eqref{G_xx_backToCartesian},
after integrating with respect to $\phi$ and multiplying by  $1/(2\pi)^2$ of the spatial inverse Fourier transform to get
\begin{eqnarray}
 \Delta \tilde{\mathcal{G}}_{xx} (q,\omega,z_0)  &=& \frac{1}{8\pi\eta} \frac{i \alpha_\mathrm{s}}{2-i\alpha_\mathrm{s} q} e^{-2q z_0} + \frac{1}{16\pi\eta} \frac{i\alpha (q z_0-1)^2}{1-i\alpha q} e^{-2q z_0},\label{OUR1}\\
 \Delta \tilde{\mathcal{G}}_{zz} (q,\omega,z_0) &=& \frac{1}{8\pi\eta} \frac{i\alpha (q z_0)^2}{1-i\alpha q} e^{-2q z_0}.\label{OUR2}
\end{eqnarray}

Hereafter, $\alpha = (\kappa_\mathrm{s}+\kappa_\mathrm{a})/6\eta \omega$ as defined in the main text.
In order to compare equations~(\ref{FF1} and \ref{FF2}) with (\ref{OUR1} and \ref{OUR2}), we note that Felderhof does not include the minus sign in the forward Fourier transform as we do in the present work. After substituting $i$ by $-i$, however, both equations have the same mathematical form, leading together with the definition of $\alpha_\mathrm{s}$ in Eq.~\eqref{defAlpha} to the identification $\kappa_\mathrm{s}/3=G_\mathrm{s}$ in agreement with earlier works  \cite{lac04}. 
Nevertheless, by using the fact that $E_\mathrm{s} = 2(1+\nu) G_\mathrm{s}$ and $C = \nu/(1-\nu)$ \cite{lac04}, we find that
\begin{equation}
 \frac{\alpha_\mathrm{k}}{\alpha} =  \frac{3+5C}{2(1+C)^2}. 
\end{equation}

Thus, for the neo-Hookean model with $C=1$ we find $\alpha=\alpha_\mathrm{k}$ and thus both models agree. However, in the general case of an arbitrary dilatation coefficient $C$ of the Skalak model, the two quantities are different, meaning that the constitutive law we use is different from the one used by Felderhof.

\subsection{Comparison to the work of Bickel \cite{Bickel_2006}}

Bickel \cite{Bickel_2006} studied the Brownian motion near a liquid-like membrane endowed with bending, but not with shear resistance. He provided the following correction to the mobility tensor
\begin{equation}
 \Delta \tilde{\mathcal{G}}_{kl} = \frac{i}{4\eta q} \frac{\omega_q}{\omega-i\omega_q} \gamma_k(q,z)\gamma(q,z_0) \mathcal{M}_{kl},
 \label{bickelEquation}
\end{equation}
where $\omega_q = \kappa_\mathrm{b}(q^4+\xi_{\parallel}^{-4})/4\eta q $ and  $\mathcal{M}_{kk} = 1$, $\mathcal{M}_{xy} = \mathcal{M}_{yx} = 1$, $\mathcal{M}_{xz} = \mathcal{M}_{yz} = -i$ and $\mathcal{M}_{zx} = \mathcal{M}_{zy} = i$.
The functions $\gamma_k$ are given by
\begin{eqnarray}
 \gamma_x(q,z) &=& q_x z e^{-q|z|},\\
 \gamma_y(q,z) &=& q_y z e^{-q|z|},\\
 \gamma_z(q,z) &=& (1+q|z|) e^{-q|z|}.
\end{eqnarray}

The correlation length $\xi_{\parallel}$ is not specified in detail in \cite{Bickel_2006}. For an infinite correlation length $\xi_{\parallel}^{-4} = 0$ holds and we recover the bending contributions of Eqs.~\eqref{GllP} and \eqref{GzzP}. Since the bending contribution is most important for the perpendicular mobility $\Delta \tilde{\mathcal{G}}_{zz}$ we repeat it here from Eq.~\eqref{GzzP}:
\begin{equation}
 \Delta \tilde{\mathcal{G}}_{zz} = \frac{1}{4\eta q} \frac{i \alpha_\mathrm{b}^3 q^3 (1+q|z|) (1+q z_0)}{1-i\alpha_\mathrm{b}^3 q^3}.
\end{equation}
The same equation can be obtained from Eq. \eqref{bickelEquation} for $k=l=z$.
The result is also recovered for the other components of the Green tensor, after using the transformation equations from the framework we employed ($l$ and $t$) to the usual Cartesian coordinates ($x$ and $y$).


\section{Long-time tails for the velocity autocorrelation functions}

By considering the integrand in Eqs. \eqref{eqn:mu_perp_s_unsteady} and \eqref{eqn:mu_perp_b_unsteady}, without multiplying by $s$ we have
\begin{equation}
 \frac{\Delta \mu_{\perp} (s)}{\mu_0} = \frac{6i}{\sigma^2} \frac{R}{z_0}  \left(   \frac{s^4 \left( e^{-s}-e^{-r} \right)^2}
 {  4(r-s)s^2-\sigma^2 \beta} +
    \frac{4s^6\left(s e^{-r}-r e^{-s}\right)^2}{r(16 s^5(r-s)-r \beta_\mathrm{b}^3 \sigma^2)} \right),\label{integrandPerp}
\end{equation}
with $r=\sqrt{s^2+i\sigma^2}$.
Note that $\beta \sim \sigma^2$ and $\beta_\mathrm{b} \sim \sigma^{2/3}$.
In the limit of low frequencies, the second terms appearing in the denominators can be dropped out.
Eq. \eqref{integrandPerp} reduces to
\begin{equation}
 \frac{\Delta \mu_{\perp} (s)}{\mu_0} = \frac{3i}{2\sigma^2} \frac{R}{z_0}  \frac{rs^2 \left( e^{-s}-e^{-r} \right)^2 + s \left(s e^{-r}-r e^{-s}\right)^2}{r(r-s)}.
\end{equation}

By substituting $s=p \epsilon$ and $\sigma^2 = iM^2 \epsilon^2$, it is easy to find after expanding in powers of $\epsilon$ that
\begin{equation}
 \frac{\Delta \mu_{\perp} (s)}{\mu_0} \approx -\frac{3}{2} \frac{R}{z_0} \frac{s}{s(r+s)},
\end{equation}
which is exactly the same equation as previously found by Felderhof \cite[Eq. (4.12)]{felderhof06}.
This leads after inverse Fourier transform to a $t^{-5/2}$ long-time tail for the velocity relaxation function as discussed in Ref. \cite{felderhof06}.
Similarly, we get for small $s$ and $\sigma$ in the parallel case
\begin{equation}
 \frac{\Delta \mu_{\parallel} (s)}{\mu_0} \approx -\frac{3}{4} \frac{R}{z_0} \frac{s+2r}{s(r+s)},
\end{equation}
as obtained by Felderhof \cite[Eq. (4.8)]{felderhof06}.

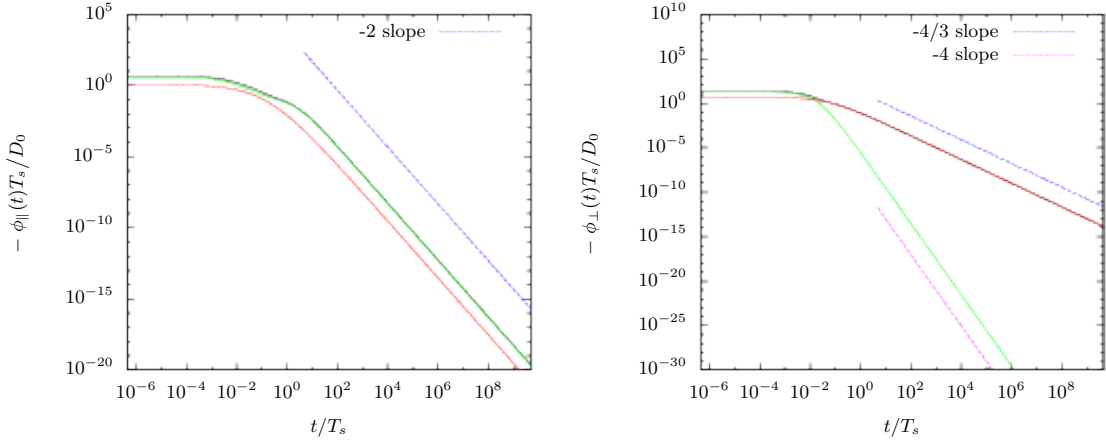
\begin{figure}
 \begin{center}
  \scalebox{0.8}{\input{VACFAll}}
 \caption{Scaled velocity autocorrelation function for the parallel (left) and the perpendicular diffusion (right).
 The shear and bending contributions are shown in green and red respectively and the total contribution is shown in black.
 The dashed lines are a guide for the eye.
 }
 \label{VACFbulkAll}
 \end{center}
\end{figure}

\subsection{Steady Stokes equations}

The velocity autocorrelation function (VACF) can be computed from the inverse Fourier transform of the particle steady mobility correction, as sated in Eq. \eqref{VACFDefinition} of the main text
\begin{equation}
 \phi_{v} (t) =  D_0 \frac{\Delta \mu (t)}{\mu_0},
\end{equation}
for $t>0$.
For the parallel motion, both the time dependent mobility correction due to shear and due to bending have a long-time tail of $t^{-2}$, as it can be seen from Eqs. \eqref{DeltaMuTimeParaShear} and \eqref{DeltaMuTimeParaBending}.
Therefore,  $\phi_{\parallel} (t)$ scales as $ t^{-2}$ at large times.

For the perpendicular case, we showed that at large times, the shear contribution has a $t^{-4}$ tail (Eq. \eqref{DeltaMuTimePerpShear}) while the bending contribution has a $t^{-4/3}$ tail (Eq. \eqref{DeltaMuTimePerpBending}).
Thus, we get a $t^{-4/3}$ tail for $\phi_{\perp} (t)$.
Fig. \ref{VACFbulkAll} illustrates the scaled velocity autocorrelation functions for both the parallel and the perpendicular diffusion.
The shear and bending contributions are also shown.


\section{Fitting formula}

We have shown in Eqs. \eqref{bendingMobilityCorrectionTimePara} and \eqref{bendingMobilityCorrectionTimePerp} that the time dependent mobility correction for the perpendicular motion and for the parallel motion respectively read
\begin{equation}
  \frac{\Delta \mu_{\parallel,\mathrm{b}} (t)}{\mu_0} = - \frac{3}{8} \frac{a}{z_0} \frac{\theta (t)}{T_b} I_{\parallel},~~~~
  \frac{\Delta {\mu}_{\perp,\mathrm{b}}(t)}{\mu_0} = -\frac{3}{4} \frac{a}{z_0} \frac{\theta(t)}{T_b} I_{\perp}
  , \label{timeDependentMobility}
\end{equation}
where
\begin{equation}
I_{\parallel} = \int_{0}^{\infty} u^5 e^{-2u-\frac{t}{T_b} u^3} d u,~~~~
I_{\perp} = \int_0^{\infty} u^3(1+u)^2 e^{-2u- \frac{t u^3}{T_b} } d  u. \label{integrals}
\end{equation}

Unfortunately, the integrals in Eq. \eqref{integrals} can not be calculated analytically.
We therefore use the following fitting formulas (Batchelor parametrization \cite{batchelor50})
\begin{equation}
 I_{\parallel} = \frac{15}{8} \frac{1}{\left(1+\tau_\mathrm{b}^{1/2} \right)^{4}},~~~~
 I_{\perp} = \frac{15}{4} \frac{1}{\left(1+\tau_\mathrm{b}^{2/3} \right)^{2}},\label{fittingFormulas}
\end{equation}
where $\tau_\mathrm{b} = t/r T_b$ with  $r=2/5$ for $I_\parallel$ and $r=4/9\pi$ for $I_\perp$.
We recall that $T_b = 4\eta z_0^3 / \kappa_\mathrm{b}$.

\begin{figure}
 \begin{center}
  \scalebox{0.8}{\input{bendingFittingInTime}}
 \caption{
 Comparison between the numerical evaluation of the integrals given by Eq. \eqref{integrals} and the fitting formulas given by Eq. \eqref{fittingFormulas}.
 At long times,  the integrals decay following a $t^{-2}$ law for $I_\parallel$, and as $t^{-4/3}$ for $I_\perp$. }
 \label{bendingInTime}
 \end{center}
\end{figure}
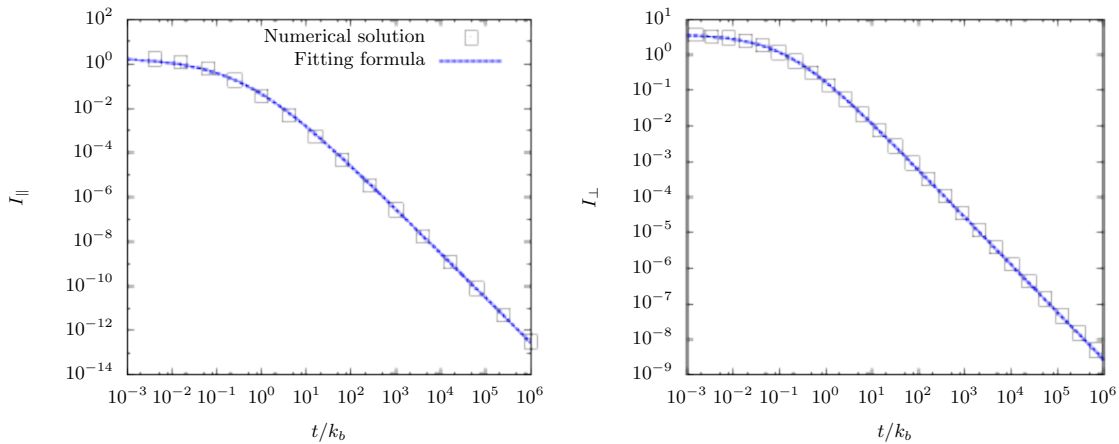

In Fig. \ref{bendingInTime}, we show a comparison between the results given by the numerical integration and the fitting formulas whose expressions are given by Eq. \eqref{fittingFormulas}.
The two are found to be in a very good agreement, suggesting that the fitting formulas can be used.

\section{Limiting case for a hard wall}

For a membrane with infinite shear and bending moduli, the hard wall limit is recovered. This is equivalent of taking the limits when $\beta$ and $\beta_\mathrm{b}$ tend to zero in the mobility correction due to shear and bending, respectively. We get
\begin{eqnarray}
 \lim_{\beta \to 0} \frac{\Delta \mu_{\parallel,\mathrm{s}} (\beta)}{\mu_0} &=& -\frac{15}{32}\frac{R}{z_0},\label{paraShearLim}\\
 \lim_{\beta_\mathrm{b}\to 0} \frac{\Delta \mu_{\parallel, \mathrm{b}} (\beta_\mathrm{b})}{\mu_0} &=& -\frac{3}{32}\frac{R}{z_0},\label{paraBenLim}\\
 \lim_{\beta\to 0} \frac{\Delta {\mu}_{\perp,\mathrm{s}} (\beta)}{\mu_0} &=& -\frac{3}{16}\frac{R}{z_0},\label{perpShearLim}\\
 \lim_{\beta_\mathrm{b}\to 0} \frac{\Delta {\mu}_{\perp,\mathrm{b}} (\beta_\mathrm{b})}{\mu_0} &=& -\frac{15}{16}\frac{R}{z_0}. \label{perpBenLim} 
\end{eqnarray}

It is worth to note here that for a membrane with infinite bending modulus, as stated in Eqs. \eqref{paraBenLim} and \eqref{perpBenLim}, the mobility correction to the first order is identical to the one corresponding to a flat interface separating two fluids with the same viscosity ratio \cite{Lee_1979, Bickel_2006}. On the other hand, when all the moduli are taken to infinity, then the total mobility correction is nothing but the first order mobility correction for a particle near a rigid wall given in the main text. 
We note that in the large membrane moduli limit the contribution due to shear is five times more significant than the one due to bending for the parallel motion. For the perpendicular motion, the contribution due to bending is five times larger than the one due to shear.


\input{biblio2}

%% file: steadyUnsteady.tex
\begingroup
  \makeatletter
  \providecommand\color[2][]{%
    \GenericError{(gnuplot) \space\space\space\@spaces}{%
      Package color not loaded in conjunction with
      terminal option `colourtext'%
    }{See the gnuplot documentation for explanation.%
    }{Either use 'blacktext' in gnuplot or load the package
      color.sty in LaTeX.}%
    \renewcommand\color[2][]{}%
  }%
  \providecommand\includegraphics[2][]{%
    \GenericError{(gnuplot) \space\space\space\@spaces}{%
      Package graphicx or graphics not loaded%
    }{See the gnuplot documentation for explanation.%
    }{The gnuplot epslatex terminal needs graphicx.sty or graphics.sty.}%
    \renewcommand\includegraphics[2][]{}%
  }%
  \providecommand\rotatebox[2]{#2}%
  \@ifundefined{ifGPcolor}{%
    \newif\ifGPcolor
    \GPcolorfalse
  }{}%
  \@ifundefined{ifGPblacktext}{%
    \newif\ifGPblacktext
    \GPblacktexttrue
  }{}%
  \let\gplgaddtomacro\g@addto@macro
  \gdef\gplbacktext{}%
  \gdef\gplfronttext{}%
  \makeatother
  \ifGPblacktext
    \def\colorrgb#1{}%
    \def\colorgray#1{}%
  \else
    \ifGPcolor
      \def\colorrgb#1{\color[rgb]{#1}}%
      \def\colorgray#1{\color[gray]{#1}}%
      \expandafter\def\csname LTw\endcsname{\color{white}}%
      \expandafter\def\csname LTb\endcsname{\color{black}}%
      \expandafter\def\csname LTa\endcsname{\color{black}}%
      \expandafter\def\csname LT0\endcsname{\color[rgb]{1,0,0}}%
      \expandafter\def\csname LT1\endcsname{\color[rgb]{0,1,0}}%
      \expandafter\def\csname LT2\endcsname{\color[rgb]{0,0,1}}%
      \expandafter\def\csname LT3\endcsname{\color[rgb]{1,0,1}}%
      \expandafter\def\csname LT4\endcsname{\color[rgb]{0,1,1}}%
      \expandafter\def\csname LT5\endcsname{\color[rgb]{1,1,0}}%
      \expandafter\def\csname LT6\endcsname{\color[rgb]{0,0,0}}%
      \expandafter\def\csname LT7\endcsname{\color[rgb]{1,0.3,0}}%
      \expandafter\def\csname LT8\endcsname{\color[rgb]{0.5,0.5,0.5}}%
    \else
      \def\colorrgb#1{\color{black}}%
      \def\colorgray#1{\color[gray]{#1}}%
      \expandafter\def\csname LTw\endcsname{\color{white}}%
      \expandafter\def\csname LTb\endcsname{\color{black}}%
      \expandafter\def\csname LTa\endcsname{\color{black}}%
      \expandafter\def\csname LT0\endcsname{\color{black}}%
      \expandafter\def\csname LT1\endcsname{\color{black}}%
      \expandafter\def\csname LT2\endcsname{\color{black}}%
      \expandafter\def\csname LT3\endcsname{\color{black}}%
      \expandafter\def\csname LT4\endcsname{\color{black}}%
      \expandafter\def\csname LT5\endcsname{\color{black}}%
      \expandafter\def\csname LT6\endcsname{\color{black}}%
      \expandafter\def\csname LT7\endcsname{\color{black}}%
      \expandafter\def\csname LT8\endcsname{\color{black}}%
    \fi
  \fi
  \setlength{\unitlength}{0.0500bp}%
  \begin{picture}(10800.00,4320.00)%
    \gplgaddtomacro\gplbacktext{%
      \csname LTb\endcsname%
      \put(946,704){\makebox(0,0)[r]{\strut{}-0.4}}%
      \put(946,1313){\makebox(0,0)[r]{\strut{}-0.3}}%
      \put(946,1923){\makebox(0,0)[r]{\strut{}-0.2}}%
      \put(946,2532){\makebox(0,0)[r]{\strut{}-0.1}}%
      \put(946,3141){\makebox(0,0)[r]{\strut{} 0}}%
      \put(946,3750){\makebox(0,0)[r]{\strut{} 0.1}}%
      \put(1078,484){\makebox(0,0){\strut{}$10^{0}$}}%
      \put(1569,484){\makebox(0,0){\strut{}$10^{1}$}}%
      \put(2059,484){\makebox(0,0){\strut{}$10^{2}$}}%
      \put(2550,484){\makebox(0,0){\strut{}$10^{3}$}}%
      \put(3041,484){\makebox(0,0){\strut{}$10^{4}$}}%
      \put(3531,484){\makebox(0,0){\strut{}$10^{5}$}}%
      \put(4022,484){\makebox(0,0){\strut{}$10^{6}$}}%
      \put(4512,484){\makebox(0,0){\strut{}$10^{7}$}}%
      \put(5003,484){\makebox(0,0){\strut{}$10^{8}$}}%
      \put(176,2379){\rotatebox{-270}{\makebox(0,0){\strut{}$\Delta \mu_{\parallel}/\mu_0$}}}%
      \put(3040,154){\makebox(0,0){\strut{}$f$[Hz]}}%
    }%
    \gplgaddtomacro\gplfronttext{%
      \csname LTb\endcsname%
      \put(4148,1508){\makebox(0,0)[r]{\strut{}Unsteady}}%
      \csname LTb\endcsname%
      \put(4148,1288){\makebox(0,0)[r]{\strut{}Steady}}%
    }%
    \gplgaddtomacro\gplbacktext{%
      \csname LTb\endcsname%
      \put(6346,704){\makebox(0,0)[r]{\strut{}-0.8}}%
      \put(6346,1374){\makebox(0,0)[r]{\strut{}-0.6}}%
      \put(6346,2044){\makebox(0,0)[r]{\strut{}-0.4}}%
      \put(6346,2715){\makebox(0,0)[r]{\strut{}-0.2}}%
      \put(6346,3385){\makebox(0,0)[r]{\strut{} 0}}%
      \put(6346,4055){\makebox(0,0)[r]{\strut{} 0.2}}%
      \put(6478,484){\makebox(0,0){\strut{}$10^{0}$}}%
      \put(6969,484){\makebox(0,0){\strut{}$10^{1}$}}%
      \put(7459,484){\makebox(0,0){\strut{}$10^{2}$}}%
      \put(7950,484){\makebox(0,0){\strut{}$10^{3}$}}%
      \put(8440,484){\makebox(0,0){\strut{}$10^{4}$}}%
      \put(8931,484){\makebox(0,0){\strut{}$10^{5}$}}%
      \put(9421,484){\makebox(0,0){\strut{}$10^{6}$}}%
      \put(9912,484){\makebox(0,0){\strut{}$10^{7}$}}%
      \put(10402,484){\makebox(0,0){\strut{}$10^{8}$}}%
      \put(5576,2379){\rotatebox{-270}{\makebox(0,0){\strut{}$\Delta \mu_{\perp}/\mu_0$}}}%
      \put(8440,154){\makebox(0,0){\strut{}$f$[Hz]}}%
    }%
    \gplgaddtomacro\gplfronttext{%
    }%
    \gplbacktext
    \put(0,0){\includegraphics{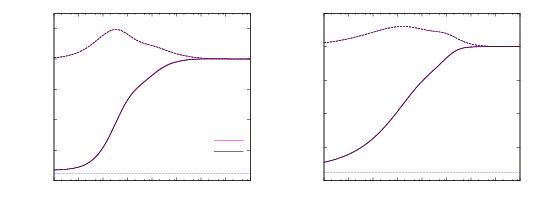}}%
    \gplfronttext
  \end{picture}%
\endgroup

%% file: VACFAll.tex
\begingroup
  \makeatletter
  \providecommand\color[2][]{%
    \GenericError{(gnuplot) \space\space\space\@spaces}{%
      Package color not loaded in conjunction with
      terminal option `colourtext'%
    }{See the gnuplot documentation for explanation.%
    }{Either use 'blacktext' in gnuplot or load the package
      color.sty in LaTeX.}%
    \renewcommand\color[2][]{}%
  }%
  \providecommand\includegraphics[2][]{%
    \GenericError{(gnuplot) \space\space\space\@spaces}{%
      Package graphicx or graphics not loaded%
    }{See the gnuplot documentation for explanation.%
    }{The gnuplot epslatex terminal needs graphicx.sty or graphics.sty.}%
    \renewcommand\includegraphics[2][]{}%
  }%
  \providecommand\rotatebox[2]{#2}%
  \@ifundefined{ifGPcolor}{%
    \newif\ifGPcolor
    \GPcolorfalse
  }{}%
  \@ifundefined{ifGPblacktext}{%
    \newif\ifGPblacktext
    \GPblacktexttrue
  }{}%
  \let\gplgaddtomacro\g@addto@macro
  \gdef\gplbacktext{}%
  \gdef\gplfronttext{}%
  \makeatother
  \ifGPblacktext
    \def\colorrgb#1{}%
    \def\colorgray#1{}%
  \else
    \ifGPcolor
      \def\colorrgb#1{\color[rgb]{#1}}%
      \def\colorgray#1{\color[gray]{#1}}%
      \expandafter\def\csname LTw\endcsname{\color{white}}%
      \expandafter\def\csname LTb\endcsname{\color{black}}%
      \expandafter\def\csname LTa\endcsname{\color{black}}%
      \expandafter\def\csname LT0\endcsname{\color[rgb]{1,0,0}}%
      \expandafter\def\csname LT1\endcsname{\color[rgb]{0,1,0}}%
      \expandafter\def\csname LT2\endcsname{\color[rgb]{0,0,1}}%
      \expandafter\def\csname LT3\endcsname{\color[rgb]{1,0,1}}%
      \expandafter\def\csname LT4\endcsname{\color[rgb]{0,1,1}}%
      \expandafter\def\csname LT5\endcsname{\color[rgb]{1,1,0}}%
      \expandafter\def\csname LT6\endcsname{\color[rgb]{0,0,0}}%
      \expandafter\def\csname LT7\endcsname{\color[rgb]{1,0.3,0}}%
      \expandafter\def\csname LT8\endcsname{\color[rgb]{0.5,0.5,0.5}}%
    \else
      \def\colorrgb#1{\color{black}}%
      \def\colorgray#1{\color[gray]{#1}}%
      \expandafter\def\csname LTw\endcsname{\color{white}}%
      \expandafter\def\csname LTb\endcsname{\color{black}}%
      \expandafter\def\csname LTa\endcsname{\color{black}}%
      \expandafter\def\csname LT0\endcsname{\color{black}}%
      \expandafter\def\csname LT1\endcsname{\color{black}}%
      \expandafter\def\csname LT2\endcsname{\color{black}}%
      \expandafter\def\csname LT3\endcsname{\color{black}}%
      \expandafter\def\csname LT4\endcsname{\color{black}}%
      \expandafter\def\csname LT5\endcsname{\color{black}}%
      \expandafter\def\csname LT6\endcsname{\color{black}}%
      \expandafter\def\csname LT7\endcsname{\color{black}}%
      \expandafter\def\csname LT8\endcsname{\color{black}}%
    \fi
  \fi
  \setlength{\unitlength}{0.0500bp}%
  \begin{picture}(10800.00,4320.00)%
    \gplgaddtomacro\gplbacktext{%
      \csname LTb\endcsname%
      \put(6478,704){\makebox(0,0)[r]{\strut{}$10^{-30}$}}%
      \put(6478,1123){\makebox(0,0)[r]{\strut{}$10^{-25}$}}%
      \put(6478,1542){\makebox(0,0)[r]{\strut{}$10^{-20}$}}%
      \put(6478,1961){\makebox(0,0)[r]{\strut{}$10^{-15}$}}%
      \put(6478,2380){\makebox(0,0)[r]{\strut{}$10^{-10}$}}%
      \put(6478,2798){\makebox(0,0)[r]{\strut{}$10^{-5}$}}%
      \put(6478,3217){\makebox(0,0)[r]{\strut{}$10^{0}$}}%
      \put(6478,3636){\makebox(0,0)[r]{\strut{}$10^{5}$}}%
      \put(6478,4055){\makebox(0,0)[r]{\strut{}$10^{10}$}}%
      \put(6691,484){\makebox(0,0){\strut{}$10^{-6}$}}%
      \put(7165,484){\makebox(0,0){\strut{}$10^{-4}$}}%
      \put(7639,484){\makebox(0,0){\strut{}$10^{-2}$}}%
      \put(8113,484){\makebox(0,0){\strut{}$10^{0}$}}%
      \put(8587,484){\makebox(0,0){\strut{}$10^{2}$}}%
      \put(9061,484){\makebox(0,0){\strut{}$10^{4}$}}%
      \put(9535,484){\makebox(0,0){\strut{}$10^{6}$}}%
      \put(10009,484){\makebox(0,0){\strut{}$10^{8}$}}%
      \put(5576,2379){\rotatebox{-270}{\makebox(0,0){\strut{}$-\left. \phi_{\perp} (t) T_s \middle/ D_0 \right.$}}}%
      \put(8506,154){\makebox(0,0){\strut{}$t/T_s$}}%
    }%
    \gplgaddtomacro\gplfronttext{%
      \csname LTb\endcsname%
      \put(9415,3882){\makebox(0,0)[r]{\strut{}-4/3 slope}}%
      \csname LTb\endcsname%
      \put(9415,3662){\makebox(0,0)[r]{\strut{}-4 slope}}%
    }%
    \gplgaddtomacro\gplbacktext{%
      \csname LTb\endcsname%
      \put(1078,704){\makebox(0,0)[r]{\strut{}$10^{-20}$}}%
      \put(1078,1374){\makebox(0,0)[r]{\strut{}$10^{-15}$}}%
      \put(1078,2044){\makebox(0,0)[r]{\strut{}$10^{-10}$}}%
      \put(1078,2715){\makebox(0,0)[r]{\strut{}$10^{-5}$}}%
      \put(1078,3385){\makebox(0,0)[r]{\strut{}$10^{0}$}}%
      \put(1078,4055){\makebox(0,0)[r]{\strut{}$10^{5}$}}%
      \put(1291,484){\makebox(0,0){\strut{}$10^{-6}$}}%
      \put(1765,484){\makebox(0,0){\strut{}$10^{-4}$}}%
      \put(2240,484){\makebox(0,0){\strut{}$10^{-2}$}}%
      \put(2714,484){\makebox(0,0){\strut{}$10^{0}$}}%
      \put(3188,484){\makebox(0,0){\strut{}$10^{2}$}}%
      \put(3662,484){\makebox(0,0){\strut{}$10^{4}$}}%
      \put(4136,484){\makebox(0,0){\strut{}$10^{6}$}}%
      \put(4610,484){\makebox(0,0){\strut{}$10^{8}$}}%
      \put(176,2379){\rotatebox{-270}{\makebox(0,0){\strut{}$-\left. \phi_{\parallel} (t) T_s \middle/ D_0 \right.$}}}%
      \put(3106,154){\makebox(0,0){\strut{}$t/T_s$}}%
    }%
    \gplgaddtomacro\gplfronttext{%
      \csname LTb\endcsname%
      \put(4016,3882){\makebox(0,0)[r]{\strut{}-2 slope}}%
    }%
    \gplbacktext
    \put(0,0){\includegraphics{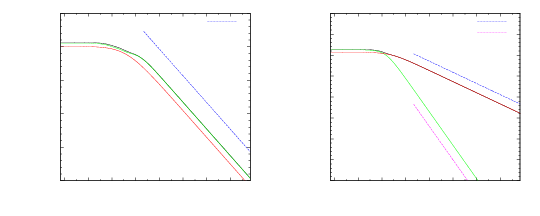}}%
    \gplfronttext
  \end{picture}%
\endgroup

%% file: bendingFittingInTime.tex
\begingroup
  \makeatletter
  \providecommand\color[2][]{%
    \GenericError{(gnuplot) \space\space\space\@spaces}{%
      Package color not loaded in conjunction with
      terminal option `colourtext'%
    }{See the gnuplot documentation for explanation.%
    }{Either use 'blacktext' in gnuplot or load the package
      color.sty in LaTeX.}%
    \renewcommand\color[2][]{}%
  }%
  \providecommand\includegraphics[2][]{%
    \GenericError{(gnuplot) \space\space\space\@spaces}{%
      Package graphicx or graphics not loaded%
    }{See the gnuplot documentation for explanation.%
    }{The gnuplot epslatex terminal needs graphicx.sty or graphics.sty.}%
    \renewcommand\includegraphics[2][]{}%
  }%
  \providecommand\rotatebox[2]{#2}%
  \@ifundefined{ifGPcolor}{%
    \newif\ifGPcolor
    \GPcolorfalse
  }{}%
  \@ifundefined{ifGPblacktext}{%
    \newif\ifGPblacktext
    \GPblacktexttrue
  }{}%
  \let\gplgaddtomacro\g@addto@macro
  \gdef\gplbacktext{}%
  \gdef\gplfronttext{}%
  \makeatother
  \ifGPblacktext
    \def\colorrgb#1{}%
    \def\colorgray#1{}%
  \else
    \ifGPcolor
      \def\colorrgb#1{\color[rgb]{#1}}%
      \def\colorgray#1{\color[gray]{#1}}%
      \expandafter\def\csname LTw\endcsname{\color{white}}%
      \expandafter\def\csname LTb\endcsname{\color{black}}%
      \expandafter\def\csname LTa\endcsname{\color{black}}%
      \expandafter\def\csname LT0\endcsname{\color[rgb]{1,0,0}}%
      \expandafter\def\csname LT1\endcsname{\color[rgb]{0,1,0}}%
      \expandafter\def\csname LT2\endcsname{\color[rgb]{0,0,1}}%
      \expandafter\def\csname LT3\endcsname{\color[rgb]{1,0,1}}%
      \expandafter\def\csname LT4\endcsname{\color[rgb]{0,1,1}}%
      \expandafter\def\csname LT5\endcsname{\color[rgb]{1,1,0}}%
      \expandafter\def\csname LT6\endcsname{\color[rgb]{0,0,0}}%
      \expandafter\def\csname LT7\endcsname{\color[rgb]{1,0.3,0}}%
      \expandafter\def\csname LT8\endcsname{\color[rgb]{0.5,0.5,0.5}}%
    \else
      \def\colorrgb#1{\color{black}}%
      \def\colorgray#1{\color[gray]{#1}}%
      \expandafter\def\csname LTw\endcsname{\color{white}}%
      \expandafter\def\csname LTb\endcsname{\color{black}}%
      \expandafter\def\csname LTa\endcsname{\color{black}}%
      \expandafter\def\csname LT0\endcsname{\color{black}}%
      \expandafter\def\csname LT1\endcsname{\color{black}}%
      \expandafter\def\csname LT2\endcsname{\color{black}}%
      \expandafter\def\csname LT3\endcsname{\color{black}}%
      \expandafter\def\csname LT4\endcsname{\color{black}}%
      \expandafter\def\csname LT5\endcsname{\color{black}}%
      \expandafter\def\csname LT6\endcsname{\color{black}}%
      \expandafter\def\csname LT7\endcsname{\color{black}}%
      \expandafter\def\csname LT8\endcsname{\color{black}}%
    \fi
  \fi
  \setlength{\unitlength}{0.0500bp}%
  \begin{picture}(10800.00,4320.00)%
    \gplgaddtomacro\gplbacktext{%
      \csname LTb\endcsname%
      \put(6346,704){\makebox(0,0)[r]{\strut{}$10^{-9}$}}%
      \put(6346,1039){\makebox(0,0)[r]{\strut{}$10^{-8}$}}%
      \put(6346,1374){\makebox(0,0)[r]{\strut{}$10^{-7}$}}%
      \put(6346,1709){\makebox(0,0)[r]{\strut{}$10^{-6}$}}%
      \put(6346,2044){\makebox(0,0)[r]{\strut{}$10^{-5}$}}%
      \put(6346,2380){\makebox(0,0)[r]{\strut{}$10^{-4}$}}%
      \put(6346,2715){\makebox(0,0)[r]{\strut{}$10^{-3}$}}%
      \put(6346,3050){\makebox(0,0)[r]{\strut{}$10^{-2}$}}%
      \put(6346,3385){\makebox(0,0)[r]{\strut{}$10^{-1}$}}%
      \put(6346,3720){\makebox(0,0)[r]{\strut{}$10^{0}$}}%
      \put(6346,4055){\makebox(0,0)[r]{\strut{}$10^{1}$}}%
      \put(6478,484){\makebox(0,0){\strut{}$10^{-3}$}}%
      \put(6914,484){\makebox(0,0){\strut{}$10^{-2}$}}%
      \put(7350,484){\makebox(0,0){\strut{}$10^{-1}$}}%
      \put(7786,484){\makebox(0,0){\strut{}$10^{0}$}}%
      \put(8222,484){\makebox(0,0){\strut{}$10^{1}$}}%
      \put(8658,484){\makebox(0,0){\strut{}$10^{2}$}}%
      \put(9094,484){\makebox(0,0){\strut{}$10^{3}$}}%
      \put(9530,484){\makebox(0,0){\strut{}$10^{4}$}}%
      \put(9966,484){\makebox(0,0){\strut{}$10^{5}$}}%
      \put(10402,484){\makebox(0,0){\strut{}$10^{6}$}}%
      \put(5576,2379){\rotatebox{-270}{\makebox(0,0){\strut{}$I_{\perp}$}}}%
      \put(8440,154){\makebox(0,0){\strut{}$t/k_b$}}%
    }%
    \gplgaddtomacro\gplfronttext{%
    }%
    \gplgaddtomacro\gplbacktext{%
      \csname LTb\endcsname%
      \put(1078,704){\makebox(0,0)[r]{\strut{}$10^{-14}$}}%
      \put(1078,1123){\makebox(0,0)[r]{\strut{}$10^{-12}$}}%
      \put(1078,1542){\makebox(0,0)[r]{\strut{}$10^{-10}$}}%
      \put(1078,1961){\makebox(0,0)[r]{\strut{}$10^{-8}$}}%
      \put(1078,2380){\makebox(0,0)[r]{\strut{}$10^{-6}$}}%
      \put(1078,2798){\makebox(0,0)[r]{\strut{}$10^{-4}$}}%
      \put(1078,3217){\makebox(0,0)[r]{\strut{}$10^{-2}$}}%
      \put(1078,3636){\makebox(0,0)[r]{\strut{}$10^{0}$}}%
      \put(1078,4055){\makebox(0,0)[r]{\strut{}$10^{2}$}}%
      \put(1210,484){\makebox(0,0){\strut{}$10^{-3}$}}%
      \put(1631,484){\makebox(0,0){\strut{}$10^{-2}$}}%
      \put(2053,484){\makebox(0,0){\strut{}$10^{-1}$}}%
      \put(2474,484){\makebox(0,0){\strut{}$10^{0}$}}%
      \put(2896,484){\makebox(0,0){\strut{}$10^{1}$}}%
      \put(3317,484){\makebox(0,0){\strut{}$10^{2}$}}%
      \put(3739,484){\makebox(0,0){\strut{}$10^{3}$}}%
      \put(4160,484){\makebox(0,0){\strut{}$10^{4}$}}%
      \put(4582,484){\makebox(0,0){\strut{}$10^{5}$}}%
      \put(5003,484){\makebox(0,0){\strut{}$10^{6}$}}%
      \put(176,2379){\rotatebox{-270}{\makebox(0,0){\strut{}$I_{\parallel}$}}}%
      \put(3106,154){\makebox(0,0){\strut{}$t/k_b$}}%
    }%
    \gplgaddtomacro\gplfronttext{%
      \csname LTb\endcsname%
      \put(4016,3882){\makebox(0,0)[r]{\strut{}Numerical solution}}%
      \csname LTb\endcsname%
      \put(4016,3662){\makebox(0,0)[r]{\strut{}Fitting formula}}%
    }%
    \gplbacktext
    \put(0,0){\includegraphics{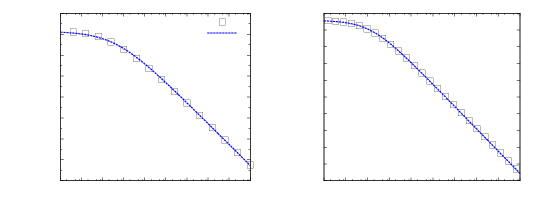}}%
    \gplfronttext
  \end{picture}%
\endgroup

%% file: biblio2.tex
%